\RequirePackage{fix-cm}
\documentclass[smallextended]{svjour3}       % onecolumn (second format)
\smartqed  % flush right qed marks, e.g. at end of proof
\usepackage{graphicx}
 \usepackage{mathptmx}      % use Times fonts if available on your TeX system
\usepackage{latexsym}

\usepackage[english]{babel}

\begin{document}

\title{Evaluation of quality measures for color quantization%\thanks{Grants or other notes
}
\subtitle{}

\author{Giuliana Ramella }
\institute{G. Ramella\at
              Institute for the Applications of Calculus, CNR\\
             Via P. Castellino 111,  80131, Naples, Italy\\
               \email{giuliana.ramella@cnr.it}           %  \\
  %         \and
     %      S. Author \at
        %      second address      
}
\maketitle
\begin{abstract}
Visual quality evaluation is one of the challenging basic problems in image processing. It also plays a central role in the shaping, implementation, optimization, and testing of many methods. The existing image quality assessment methods focused on images corrupted by common degradation types while little attention was paid to color quantization. This in spite there is a wide range of applications requiring color quantization assessment being used as a preprocessing step when color-based tasks are more efficiently accomplished on a reduced number of colors. In this paper, we propose and carry-out a quantitative performance evaluation of nine well-known and commonly used full-reference image quality assessment measures. The evaluation is done by using two publicly available and subjectively rated image quality databases for color quantization degradation and by considering suitable combinations or subparts of them. The results indicate the quality measures that have closer performances in terms of their correlation to the subjective human rating and show that the evaluation of the statistical performance of the quality measures for color quantization is significantly impacted by the selected image quality database while maintaining a similar trend on each database. The detected strong similarity both on individual databases and on databases obtained by integration provides the ability to validate the integration process and to consider the quantitative performance evaluation on each database as an indicator for performance on the other databases. The experimental results are useful to address the choice of suitable quality measures for color quantization and to improve their future employment.
\end{abstract}
\section{Introduction}
\label{intro}
The goal of Image Quality Assessment (IQA) research is to design methods for objective evaluation of quality in a way consistent with subjective human judgment. It is widely agreed that the Human Visual System (HVS) can appreciate quickly the quality of an image even if its original version is absent, which suggests that it is probably based on a high-level interpretation of the visual data and a lot of knowledge about the scene at hand.\\
IQA is an important component in most modern image processing systems such as those for acquisition, compression, restoration, enhancement, quantization, and so on. Generally speaking, IQA has at least three kinds of applications. The first application is to monitor image quality by control systems. For instance, a network server to control image streaming can use it to examine the quality of the digital image transmitted on the network and to adjust automatically the image so obtain the best quality. The second application is to optimize the methods and parameter settings. For instance, in a visual communication system, IQA can help the optimal design of each algorithm for the encoder and decoder. The third application is to benchmark image processing systems and methods. For instance, IQA can help to select the image with the best quality from multiple image processing systems for a specific task.\\
Although the knowledge about human perception mechanisms is still very limited and the existing models of the HVS generally perform under very restrictive conditions, due to this actual emergence of many new applications that require automatic real-time image quality estimate, the field of IQA is rapidly advancing. Indeed, many efforts have been directed to the design of IQA models and great progress has been achieved in this area in the past decade \cite{1,2,3}.\\
Traditionally, IQA is roughly classified into two categories: Subjective and Objective \cite{4}. For Subjective IQA, psychophysical experiments are used on different human subjects in a controlled environment and collection of human opinions are converted to a numerical value in the form of Mean Opinion Score (MOS) or Differential Mean Opinion Score (DMOS). Objective IQA includes the development of methods to perform the IQA task which is broadly classified into the following three categories based on the requirement of reference images: 1) Full-Reference image quality assessment techniques (FR-IQA); 2) Reduced-Reference image quality assessment techniques (RR-IQA); 3) No-Reference image quality assessment techniques (NR-IQA)  \cite{5,6}. As better motivated in section 2.2, we focus our attention on the FR-IQA category.\\
The IQA methods consider some specific types of distortion having various statistical properties. In most cases, IQA studies mainly focus on images corrupted by single distortions since it is an extremely difficult task to develop a method suitable to all distortions which are characterized by different attributes. In particular, very limited efforts have been devoted to the field of IQA for Color Quantization, in the following called simply CQ, which is a technique of reducing the total number of unique colors in the image \cite{7} often used as a preprocessing step for many applications which are more efficiently carried out on a small set of colors. Sometimes to CQ distortion is added dithering which is a type of noise inserted to mask the effects of quantization in the image. More details are given in section 2.3. \\
One of the most common and widely used full-reference quality measures is the Peak Signal to Noise Ratio (PSNR) \cite{8} because of its computational simplicity and clear physical meaning. However, it does not constantly well correlate with the human judgment of quality \cite{2}. Through the years, new and advanced measures like Structural Similarity Index Measure (SSIM) \cite{9} which compares luminance, contrast, and structural similarities of the original image with its distorted version has been proposed. Successively many metrics have been devised \cite{10,11,12,13,14,15,16,17,18}. Indeed, many interesting metrics based on color and/or HVS properties \cite{12,13,14,19,20} have been proposed and seem to be more appropriate for the CQ performance evaluation. However, most CQ methods, including the more recent ones  \cite{21,22}, mainly use quality measures that are not based on color or HVS properties. In particular, PSNR and SSIM and, to a lesser extent, other common quality measures, as MSSIM, VSNR, WSNR (see section 2.4), remain widely used. This phenomenon is not limited to CQ but also happens for other tasks in image analysis such as image enhancement and image scaling. This is one of the two main reasons why we focus our attention on the nine most popular quality measures.\\
Alongside the study of IQA methods, considerable effort is devoted to the development of new image databases to evaluate the performance of IQA methods \cite{2}. Many IQA databases have been established and made public \cite{23}. These databases are annotated with subjective ratings and constitute the ground truth for IQA training, testing, and benchmarking methods. The majority of these databases consider common distortions, i.e. JPEG compression, Gaussian blur, Gaussian white noise, rather than color-specific distortions \cite{24}. In particular, the distortion introduced by the CQ process has been considered initially in Tampere Image Database, namely TID2008  \cite{25}. Successively the CQ distortions have been considered in TID2013 database \cite{26}, which  is the largest so far available image database, and in Color Quantization Database, here indicated as CQD \cite{27} (see section 2.5, 3.1, 3.2). Moreover, in IQA databases, quality measures based on color/HVS properties are not publicly available included in those where CQ distortion is considered. This is the second good reason why we focus our attention on the nine quality measures. See section 3 for more detail.\\
To evaluate the performance of IQA metrics, the standard approach in the literature is to first build databases of images with various content and distortions and then collect subjective evaluation scores in the form of MOS or DMOS for all images. An alternative approach, called Maximum Differentiation (MAD) competition \cite{28}, selects automatically subsets of image pairs  to optimally distinguish the metric. Although the MAD approach allows, in some respects, to discriminate and study better the available metrics, we will refer to the standard method because this approach has never been applied to the distortion introduced by CQ and therefore there are no available data sets to apply this methodology.\\
The lack of an adequate evaluation of quality measures for CQ and, at the same time, the existence of limited data sets and quality measures for the study of the distortion introduced by CQ require consequently a close analysis of these issues. Since proposing a new CQ database or a new quality measure suitable for CQ performance evaluation is not part of our purposes, the main scope of this work is to evaluate the most popular quality measures from available data and human evaluations in the presence of distortions due to a CQ process with or without successive dithering. In particular, we want not only to conduct a quantitative performance evaluation of the quality measures for CQ but also to propose a new evaluation process based on recommendations for best practices (see section 4 for more details). Here, we test the nine well-known full-reference quality measures taken into account and we focus our attention on TID2013 and CQD databases. The test images have been selected to include the CQ distortion and a fusion of these databases is also considered (see sections 4.3 and 4.4).  The performance evaluation of these metrics (see section 5) is mainly based on the Kendall Rank Order Correlation Coefficient (KROCC) and Spearman Rank Order Correlation Coefficient (SROCC), according to VQEG \cite{29,30}, JPEG AIC \cite{31,32,33} and ITU recommendations \cite{34} (see section 2.1). Note that the fused database increases the number and heterogeneity of data taken into account since 1100 images from different data sources are considered. As a consequence, the image sample is closer to a real sample and the performance evaluation is statistically significant  (see also section 4.2).\\
From the experimental results, it is inferred: a) the quality measures that have closer performances in terms of their correlation to the subjective human rating; b) the evaluation of the statistical performance of the quality measures for color quantization is significantly impacted by the selected image quality database while maintaining a similar trend on each database; c)  the database integration process is validated and the quantitative performance evaluation on each database is an indicator for performance on the other databases.\\
Therefore, the contribution of the paper can be summarized as follows.\\
- We propose a new evaluation process in the presence of CQ distortion of the nine most popular quality measures.\\
- We perform the evaluation process on the available datasets extracted by TID2013 and CQD and on a suitable integrated version of these datasets such that to enlarge the cardinality and heterogeneity of the global database.\\
- We study and compare the experimental results in statistical terms by considering databases and their subparts obtained in different conditions.\\
- We address the problem of the choice of suitable quality measures for CQ while highlighting their features.\\
- We suggest some indications for future employment and the development of an objective quality measure for CQ.\\
The paper is organized as follows. In section 2, related notions and work, here summarily indicated, are described in more detail. In particular, recommendations by technical groups, IQA classification, the considered nine full reference quality measures, color quantization distortion, IQA Databases are outlined respectively in sub-sections 2.1-2.5. In section 3, the concerned databases from TID2013 and CQD are described. In section 4  and  section 5 respectively the motivations, objectives, methodology, and the evaluation process are discussed and illustrated also by graphical results. Discussion and conclusions are given respectively in section 6 and section 7.
\section{Related notions and work}
\label{rnw}
\subsection{Recommandations by technical groups}
\label{sec2.1}
Owing to the abundance of IQA methods with extensions and applications to other disciplines \cite{1,2,3}, it is appropriate to consider a set of rules/methods allowing us to assess the efficiency of a given IQA method. In this direction, technical groups such as VQEG (Video Quality Experts Group) or JPEG AIC (Advanced Image Coding) have focused their interest on the definition of test-plans to measure the impact of IQA metric \cite{29,30}. Currently, the work of JPEG AIC, having the aim to standardize future emerging technologies for image compression \cite{31}, resulted in two standards which provide also a set of evaluation tools that allow multiple features of such codecs to be tested and compared \cite{32} and a set of evaluation procedures  \cite{33}, similar to how the ITU-R BT.500-11 standard  \cite{34}  did for the subjective quality evaluation with television standards. This activity is a work in progress, continuously updated to arrange new coding architectures and evaluation methodologies. It is also the reason why IQA is considered as an important and compelling problem from which it is not possible to ignore. In this paper, we consider and highlight these recommendations.
\subsection{IQA classification}
\label{sec2.2}
As mentioned above, objective IQA methods are broadly classified into three categories: FR-IQA; RR-IQA; NR-IQA. The category of FR-IQA methods requires as input an original image and a processed (usually distorted) version of that image, and produce as output a quantitative assessment of the visual quality of the distorted image. The effectiveness of a FR-IQA method is assessed by examining how much the method is in accordance with human-supplied quality assessments obtained through subjective testing \cite{35}. The class of RR-IQA methods also requires a reference image but, instead of the full image information, only certain features of the reference image which possess image quality information are used for performing the IQA task of the target image \cite{36,37}. The category of NR-IQA methods performs an extremely difficult task: quality assessment of images blindly without the requirement of a reference image. Approaches based on machine learning have been developed in the literature to solve this problem to some extent although it remains mostly an open problem  \cite{38}. All three types of IQA methods can perform quite well in predicting quality. At present, FR-IQA methods have the best performing since they can generate estimates of quality that correlate highly with human ratings of quality while the research in NR-IQA and RR-IQA is much less mature even if recent methods have been shown to yield correlation coefficients that rival the most competitive FR-IQA methods.\\
Since the human visual system is easily able to identify the distortion due to CQ and/or dithering but is difficult for a method to assess the quality of CQ-distorted images without a reference image  \cite{4}, in this paper our discussion is confined to FR-IQA methods since, although extensively studied in the literature, there are still unsatisfactory conclusions regarding the assessment of this IQA class of methods.
\subsection{Color quantization distortion}
\label{sec2.3}
As above mentioned, due to constraints in bandwidth, most parts of the IQA methods in the literature are developed for assessing the quality of images by JPEG \cite{39}  and JPEG2000 \cite{40} compression technique which sometimes introduces artifacts so reducing the quality of the compressed images. Besides the compression artifacts, other common distortions are additive noise, blur, and image sharpness. Usually, distortions related to color information are disregarded or underestimated, although color images provide more relevant information to the observers rather than grayscale ones and, consequently, there is a wide range of applications requiring image color difference assessment. In addition to this lack of due attention to color-specific distortions is the fact that the quality assessment of a color image is often performed by assessing its luminance component only or its grayscale conversion. As a result, color information in the image is often largely ignored and the precision of assessment results to be influenced accordingly \cite{24}. In particular, although the CQ process is considered a fundamental process for color image analysis and a significant amount of research has been done on CQ, as mentioned above, IQA for CQ degradation has received little attention.\\
Indeed, CQ \cite{41,42,43,44,45,46,47,48,49,50} is an important step in compression methods as improper quantization can produce distortion so reducing the visual quality of the image. In the past, due to the limitations of the display hardware and to the bandwidth restrictions of computer networks, the main applications of CQ were image display \cite{42,51} and image compression \cite{52,53}. Currently, although these limitations are becoming less restrictive, CQ still maintains its practical value in these fields and has had a considerable evolution. The need of performing a CQ process frequently arises in many applications since a large number of colors (up to 16 million different colors) of a full-color image makes it difficult to handle a variety of color-based tasks which are more efficiently carried out on a reduced number of colors. Thus, CQ is considered as a prerequisite for many image processing tasks such as color segmentation \cite{54,55}, color texture analysis \cite{56}, content-based retrieval \cite{57} and also has a wide range of applicative fields. For instance, dermoscopy is an applicative field where CQ plays an important role since the colors of melanin, the most important chromophore in melanocytic neoplasms not visible by the naked eye, essentially depend on its localization in the skin \cite{58,59}.\\
Although CQ distortion has proved to be extremely important over the years and many CQ methods have been proposed, we think that the quality measures for CQ has not yet been any rigorous investigation into the performance comparison of the existing quality measures and that the available experimental data regarding CQ are rather limited. This paper is an attempt to fill this gap, at least partially.
\subsection{Full Reference quality measures}
\label{sec2.4}
Numerous full reference objective quality measures have been proposed to estimate perceived quality in visual data. These can be classified into two board types: Signal Fidelity Measures (SFM), and Perceptual Visual Quality Measures (PVQM)  \cite{60}. 
\subsubsection{The Signal Fidelity Measures }
\label{sec2.3.1}
The Signal Fidelity Measures refer to the traditional MAE (Mean Absolute Error), MSE (Mean Square Error), SNR (Signal-to-Noise Ratio), PSNR (Peak SNR), or similar  \cite{8}. Let $f(i,j)$ and $g(i,j)$  be the monochrome original and test image signal respectively $(i\leq i\leq M,\, 1\leq j \leq N):$
\begin{equation}
MSE= \displaystyle\frac{1}{MN}\sum_{i=1}^{M}\sum_{j=1}^{N} \vert \vert f(i,j)-g(i,j)\vert \vert^2
\end{equation}
\begin{equation}
PSNR=20 \displaystyle \log_{10}\left( \frac{\max_f}{\sqrt{MSE}}\right)
\end{equation}
\begin{equation}
SNR=10 \displaystyle \log_{10}\left( \frac{P_f}{P_g}\right)
\end{equation}
where $\max_f$ is the maximum possible value of the image pixel. When the pixels are represented using 8 bits per sample $\max_f$ is equal to 255. More generally, when samples are represented using linear PCM with B bits per sample, $\max_f$ is $2^{B}$-1. There are two different ways to calculate the PSNR of a color image with three RGB values per pixel. The first way exploits the property of the human eye which is very sensitive to luma information and computes the PSNR for color images by converting the image to a color space, such as YCbCr, by separating the intensity Y (luma) channel which represents a weighted average of R, G, and B and by computing the PSNR only for the luma component according to (1). The second way is based on the definition of PSNR given by (1) except the MSE is the sum over all squared value differences divided by image size and by three. In equation (3), $P_f$ and $P_g$ are respectively the power of the signal $f(i,j)$ and the signal $g(i,j)$. Since in TID2013 the first computation of PSNR is employed, we prefer to use it in the following.
Although these measures are simple, well defined, with clear physical meanings, reflect picture quality change, and widely accepted, they can be result not suitable for predicting perceived visual quality, especially when non-additive noise is present \cite{61}. In particular, PSNR has been shown to perform poorly when it comes to estimating the quality of images as perceived by humans  \cite{62}.
\subsubsection{The Perceptual Visual Quality Measures}
\label{sec2.3.2}
The Perceptual Visual Quality Measures take advantage of the known characteristics of HVS and measure image quality by estimating perceived errors. To PVQM category belong the following measures: UQI, SSIM, VIFP, VSNR, NQM, WSNR.

The Universal Quality Index (UQI) models any image distortion via a combination of three factors or components: loss of correlation, luminance distortion, and contrast distortion  \cite{63}. 
\begin{equation}
UQI= \displaystyle \frac{\sigma_{fg}}{\sigma_f\sigma_g}  \frac{2 \bar f \bar g}{(\bar f)^2+(\bar g)^2} \frac{2 \sigma_{f}\sigma_{g}}{\sigma_f^2+\sigma_g^2} 
\end{equation}
where
\[
\bar f= \displaystyle\frac{1}{MN}\sum_{i=1}^{M}\sum_{j=1}^{N} f(i,j) \qquad \qquad  \bar g= \displaystyle\frac{1}{MN}\sum_{i=1}^{M}\sum_{j=1}^{N} g(i,j) 
\]
\[
\sigma_{fg} = \displaystyle\frac{1}{M+N-1}\sum_{i=1}^{M}\sum_{j=1}^{N}( f(i,j)-\bar f)(g(i,j)-\bar g)
\]
\[
\sigma_{f}^2 = \displaystyle\frac{1}{M+N-1}\sum_{i=1}^{M}\sum_{j=1}^{N}( f(i,j)-\bar f)^2
\]
\[
\sigma_{g}^2 = \displaystyle\frac{1}{M+N-1}\sum_{i=1}^{M}\sum_{j=1}^{N}( g(i,j)-\bar g)^2
\]
In (4) the first component is the correlation coefficient, which measures the degree of linear correlation between images $f$ and $g$. It varies in the range [-1, 1]. The best value 1 is obtained when $f$ and $g$ are linearly related, which means that  $g(i,j)$ =$a$ $f(i,j)$ +$ b$ for all possible values of $i$ and $j$. The second component varies in [0, 1] and measures the distance between the average luminance of the images. Since $\sigma_{f}$ and $\sigma_{g}$ can be considered as estimates of the contrast of $f$ and $g$, the third component measures how similar the contrasts of the images are. The value range for this component is also [0,1]. The range of values for UQI is [-1, 1]. The best value 1 is achieved if and only if the images are identical.\\
The Structural Similarity index (SSIM) \cite{9}  is calculated on various windows of an image. This quality measure between two windows x and y of common size NxN is: 
\begin{equation}
SSIM(x,y)= \displaystyle \frac{(2 \mu_x\mu_y+c_1)(2 \sigma_{xy}+c_2)}{(\mu_x^2+\mu_y^2+c_1)( \sigma_x^2+\sigma_{y}^2+c_2)} 
\end{equation}
where $\mu_x$ and $\mu_y$ are respectively the average of $x$ and of $y$, $\sigma_x$ and $\sigma_y$ are respectively the variance of $x$ and of $y$, $\sigma_{xy}$ the correlation coefficient of $x$ and $y$, with 
\[
c_1=(k_1 L)^2
\]
\[
c_2=(k_2 L)^2
\]
\[L=2^{B}-1\] 
and with: $k_1$=$0.01$, $k_2$=$0.03$ by default.\\
As stated in \cite{9}, SSIM is applied to the local region using a sliding window approach. Starting from the top-left corner of the image, a sliding window of size 8x8 moves pixel by pixel horizontally and vertically through all the row and column of the image until the bottom-right corner is reached. The overall image quality is obtained by computing the average of SSIM values over all the windows.\\
The principal idea underlying the definition of SSIM is that the HVS is highly adapted to extract structural information from visual scenes. Indeed, from equation (5), SSIM can be considered as the product of three parts: luminance comparison, $l(x,y)$, contrast comparison, $c(x,y)$, and structure comparison, $s(x,y)$:
\[
l(x,y)= \displaystyle \frac{2 \mu_x\mu_y+c_1}{\mu_x^2+\mu_y^2+c_1} 
\]
\[c(x,y)= \displaystyle \frac{2 \sigma_x\sigma_y+c_2}{\sigma_x^2+\sigma_y^2+c_2} 
\]
\[
s(x,y)= \displaystyle \frac{2 \sigma_{xy}+\frac{c_2}{2}}{\sigma_x\sigma_y+\frac{c_2}{2}} 
\]
If $M$ is the total number of windows is possible to compute the Mean Structural SIMilarity index (MSSIM), given as \cite{64}:
\begin{equation}
MSSIM = \displaystyle \frac{1}{M} \sum_{j=1}^{M} SSIM(x_j,y_j)
\end{equation}
The visual information fidelity (VIFP) in pixel domain is derived from quantification of two mutual information quantities: the mutual information between the input and the output of the HVS channel when no distortion channel is present and the mutual information between the input of the distortion channel and the output of the HVS channel for the test image \cite{10}. This definition is based on the assumption that a) the reference signal has passed through another distortion channel before entering the HVS; b) combining these two quantities, a visual information fidelity measure for IQA can be derived. However, in presence of distorition significantly different from blur and white noise, such as JPEG compression, the model fails to reproduce the perceptual annoyance adequately. In this case, some assumptions are needed about the source, distortion, and HVS models  to implement VIFP criterion.\\
Another image quality metric with the more theoretical ground is the VSNR  \cite{11} which operates in two stages. In the first stage, the contrast threshold for distortion detection is computed via wavelet-based models of visual masking and visual summation, to determine the visible distortion in the test image. If the detected distortion is below the threshold of detection, the test image is considered to be of perfect visual fidelity (VSNR = infinity). If the detected distortion is above the threshold, the property of perceived contrast and the mid-level visual property of global precedence, modeled as Euclidean distances in distortion-contrast space of a multiscale wavelet decomposition, are computed. In this case, VSNR is computed based on a simple linear sum of these distances. See \cite{11} for more detail.\\
In the Noise Quality Measure (NQM), based on Peli's contrast pyramid  \cite{65}, a degraded image is modeled as an original image subjective to linear frequency distortion and additive noise injection  \cite{66}. These two sources of degradation are considered independent and are decoupled into two quality measures: a distortion measure (DM) of the effect of frequency distortion, and a noise quality measure (NQM) of the effect of additive noise. NQM takes into account the following: 1) variation in contrast sensitivity with distance, image dimensions, and spatial frequency; 2) variation in the local mean luminance; 3) contrast interaction between spatial frequencies; 4) contrast masking effects. \\
For an image of size MxN, weighted Signal to Noise Ratio (WSNR) is defined as \cite{67}:
\begin{equation}
WSNR= \displaystyle 10 \log_{10}\left(  \frac{ \sum_{u,v} \vert X(u,v) C(u,v)\vert^2}{ \sum_{u,v}  \vert X(u,v) - Y(u,v)C(u,v) \vert^2 }  \right)
\end{equation}
where $X(u,v)$, $Y(u,v)$ and $C(u,v)$ represent the Discrete Fourier Transforms (DFT’s) of the input image and CSF, respectively, and $i\leq u \leq M$ and $\ \leq v \leq N. $\\
WSNR simulates the HVS properties by filtering both the reference and distorted images with Contrast Sensitivity Function (CSF) and then compute the SNR. \\
The aforementioned quality measures don’t make use directly of color information which, on the contrary, simplifies the identification and extraction of objects in a scene. They are designed specifically for grayscale images and are extended to incorporate color images, by applying these metrics to each of three RGB color channels individually, and then combining the quality score for each channel together. As a consequence, this approach is related to human perception to a small extent, mainly because RGB color space is not able to represent one color as it is perceived by HVS.\\
Since such naive extension is suboptimal, quality measures based on color perception have been proposed \cite{12,13,14,19,20}  and the role of HVS also considered in recently proposed quality measures\cite{2} such as HVS-inspired \cite{15}, bio-inspired \cite{16}, based on saliency \cite{17}  or a suitable combination of different perceptual factors \cite{18}. In particular, in the color community, the standard metric used to judge how well an image represents perceived color to a reference image is the color-difference formula, often indicated by Delta E [42]. \\
Another evident limitation of these classical nine quality measures is that the considered type of distortion is additive noise \cite{66}, blurring \cite{68}, image sharpness \cite{4,69,70,71} (see also section 2.4), while the number of quality measures for color images dedicated to CQ degradation is rather limited \cite{72} and only recently the full reference quality measures dedicated to CQ \cite{73,74,75} have been proposed. Still, recently specific image quality measures for other types of degradation have been proposed, i. e. for contrast degradation \cite{76}, for blurring \cite{77}. Unfortunately, although there is a wide choice of suitable quality measures for color processing and specifically defined for CQ, since a) most CQ methods including recent ones, do not use such metrics; b) data about these new recent CQ quality measures are not publicly available; it is not possible to carry out a comparative assessment of them at the present.
\subsection{IQA Databases}
\label{sec2.5}
By excluding the alternative MAD approach \cite{28}, to evaluate the performance of IQA metrics, the standard approach in the literature is to first build databases of images with various content and distortions and then collect subjective evaluation scores for all images. Thus, various IQA databases have been established and made public. Well-known subject-rated image databases extensively employed in the training and testing processes in the development and benchmarking of a majority of state-of-the-art IQA models include: LIVE \cite{78}, IVC \cite{79}, TID2008 \cite{25}, MICT \cite{80}, CQSIQ \cite{81}, VCL \cite{82}], TID2013 \cite{26}] and MDIQ \cite{23}. We remark that the most common distortion types shared by these databases are JPEG compression, JPEG2000 compression, white Gaussian noise contamination, and Gaussian blur. Usually, the existing currently available databases include more than a distortion on which often the quality assessment is generally carried out globally by considering as equivalent different types of noise. Some exceptions are, for instance, the database proposed in \cite{83} for blurring distortion and the aforementioned CQD database \cite{27}.\\
The specific subjective testing methodologies vary, but each image in the databases is labeled with a MOS, which represents the average subjective opinion about the quality of the image and is often referred to as the “ground truth” quality score of the image. The typical size of these databases is in the order of hundreds or a few thousand images. LIVE is one of the most widely used databases in IQA research, it has 779 distorted images with 5 distortion types and 5 distortion levels. TID2013 is also widely used to evaluate IQA metrics, with 3000 distorted images with 24 types and 5 levels of distortions. The database MDIQ contains 1600 distorted images, each obtained by multiple distortions of random types and levels. For more details regarding IQA databases see \cite{23,84}. \\
These databases have reached the main goals for which they are created (as well as other supplementary goals) well enough. In particular, the aforementioned databases have allowed comparing performance for different full-reference visual quality metrics. The achieved performances have been utilized for a better understanding of the drawbacks of the analyzed metrics and have stimulated the design of new metrics and/or improvement (modification) of existing ones. However, the obtained results do not allow to make a proper choice of metrics suitable for a particular application in the best way and, as mentioned above, one of the most important challenges for IQA is the absence of a suitable database. Moreover, the considered degradation type is often not enough to study specific phenomena. In particular, the available databases are not enough in number and type to study the image quality related to color and in particular, that related to CQ.
\section{Test Data for CQ}
\label{sec3}
In this paper the test data which we employ are selected from two publicly available image quality databases: TID2013 and CQD to include only the CQ degradations. In this section, a brief description of these two databases, functional to the discussion of the evaluation process described in sections 4 and 5, is given. For more details regarding these databases see \cite{26} and \cite{27}, respectively.
\subsection{Tampere Image Database (TID2013)}
\label{sec3.1}
\begin{figure}
\begin{center}
  \includegraphics [width=0.5\textwidth]{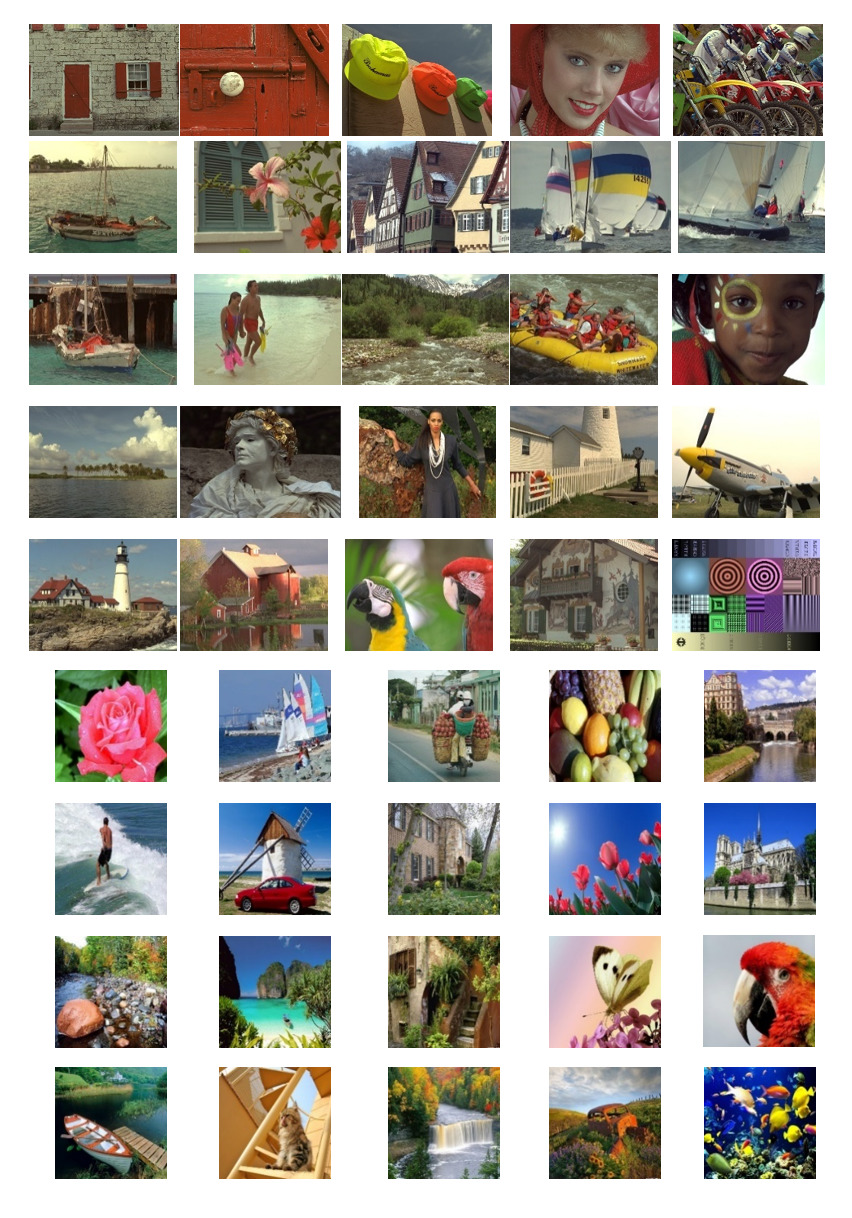}
\caption{Reference images of TID2013 rows 1-5; reference images of CQD rows 6-10.}
\label{fig:1}       % Give a unique label
\end{center}
\end{figure}

\begin{figure}
\begin{center}
 \includegraphics[width=0.23\textwidth]{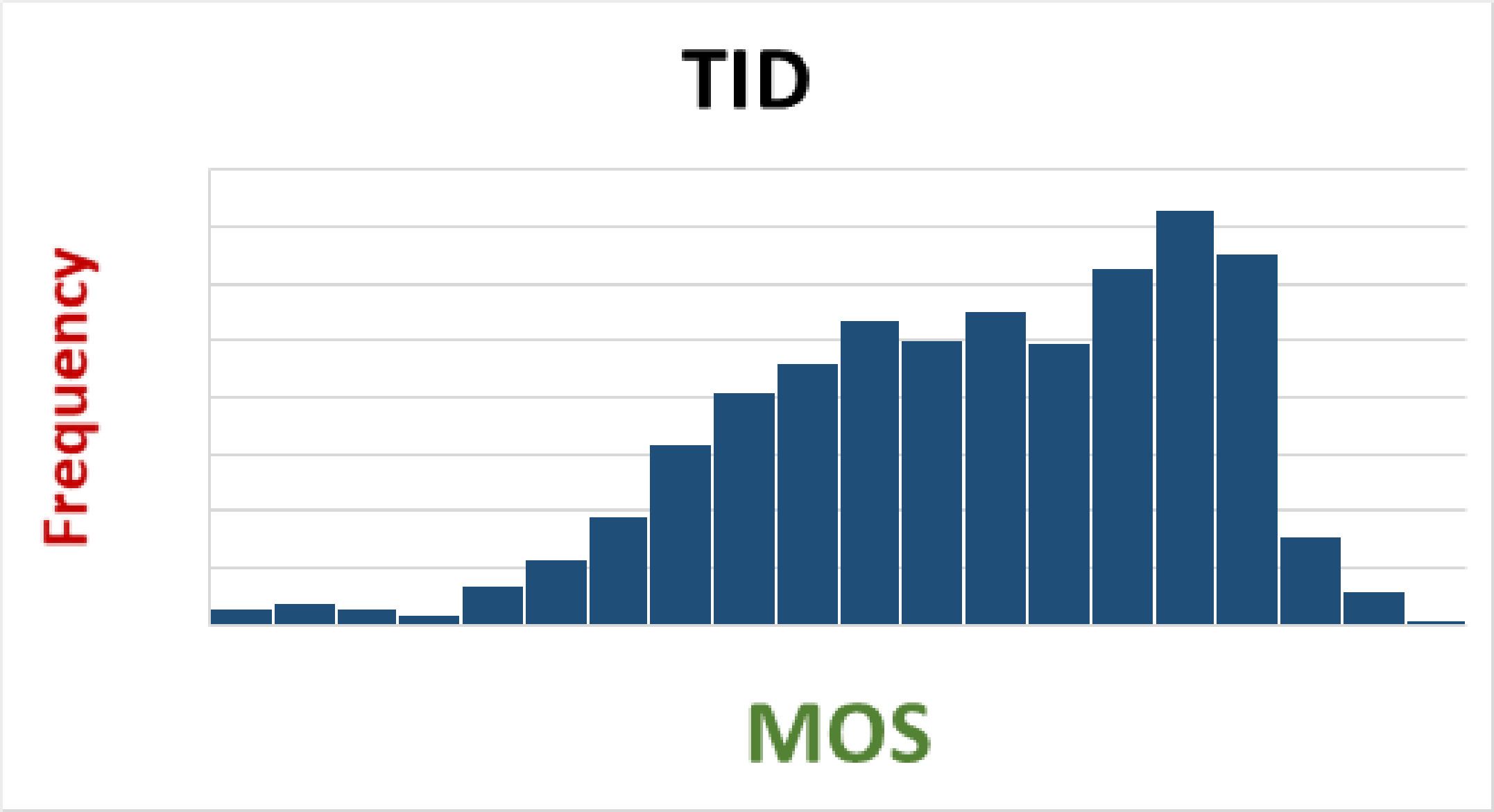}
  \includegraphics[width=0.23\textwidth]{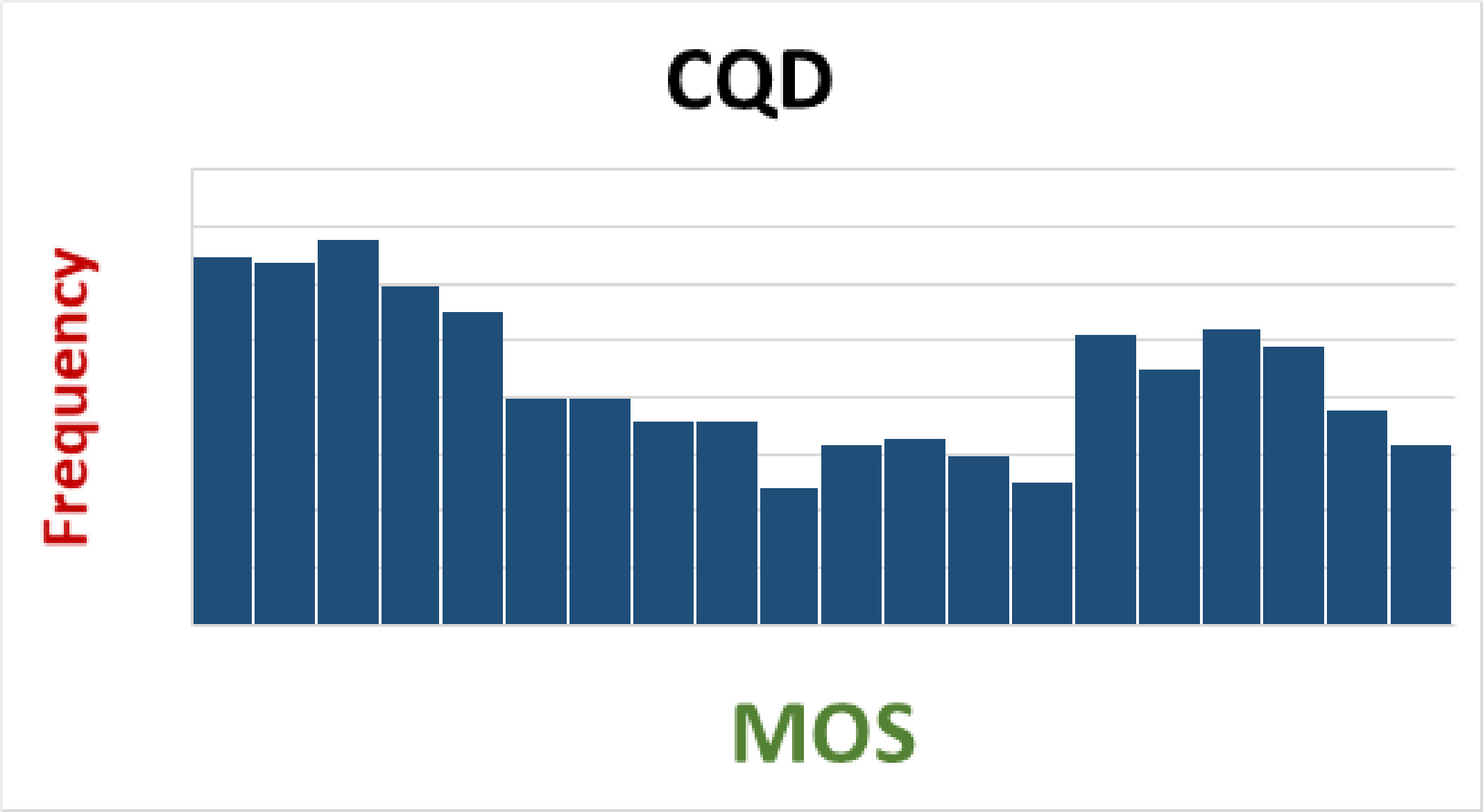}
 \includegraphics[width=0.23\textwidth]{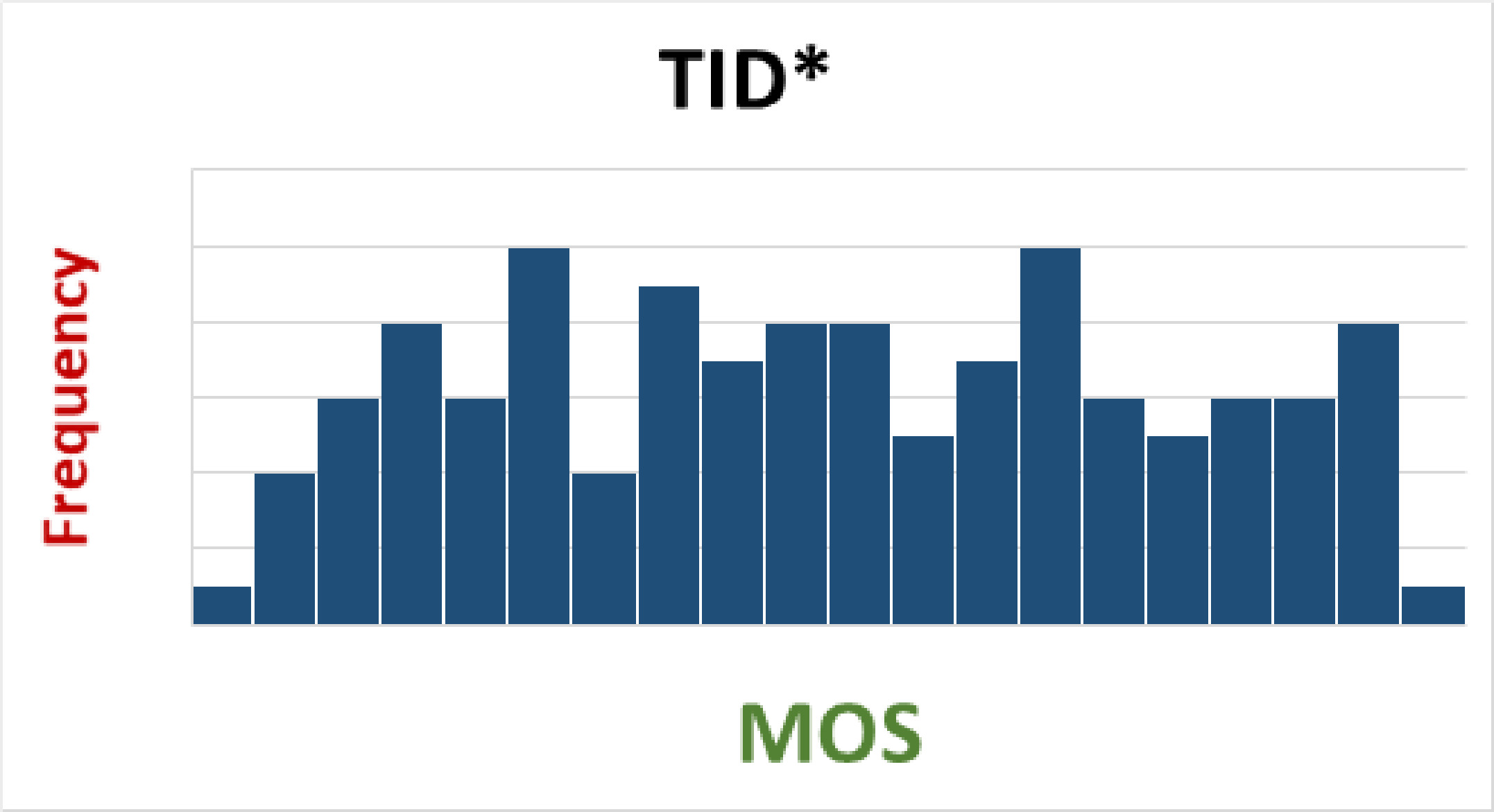}
  \includegraphics[width=0.23\textwidth]{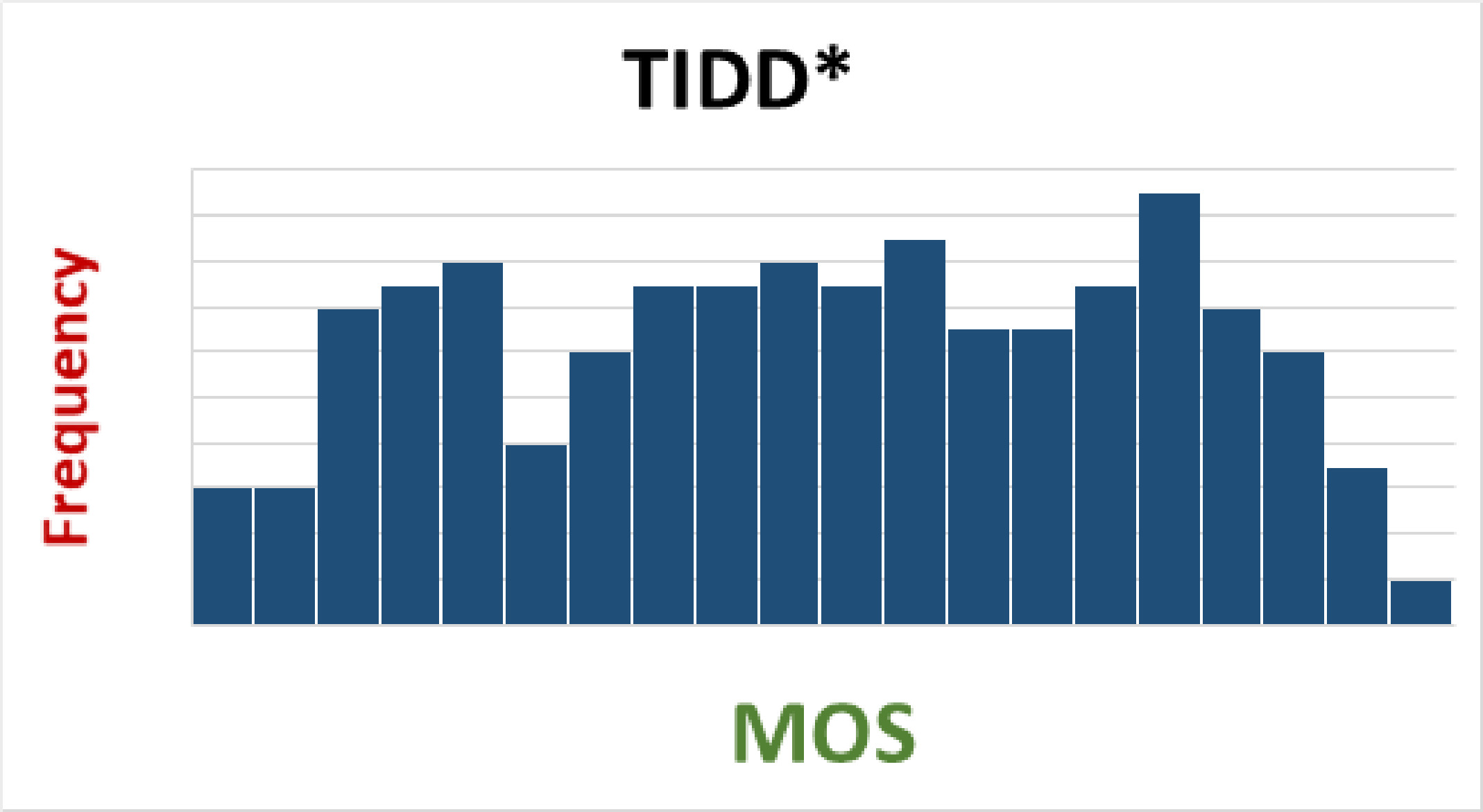}
 \includegraphics[width=0.23\textwidth]{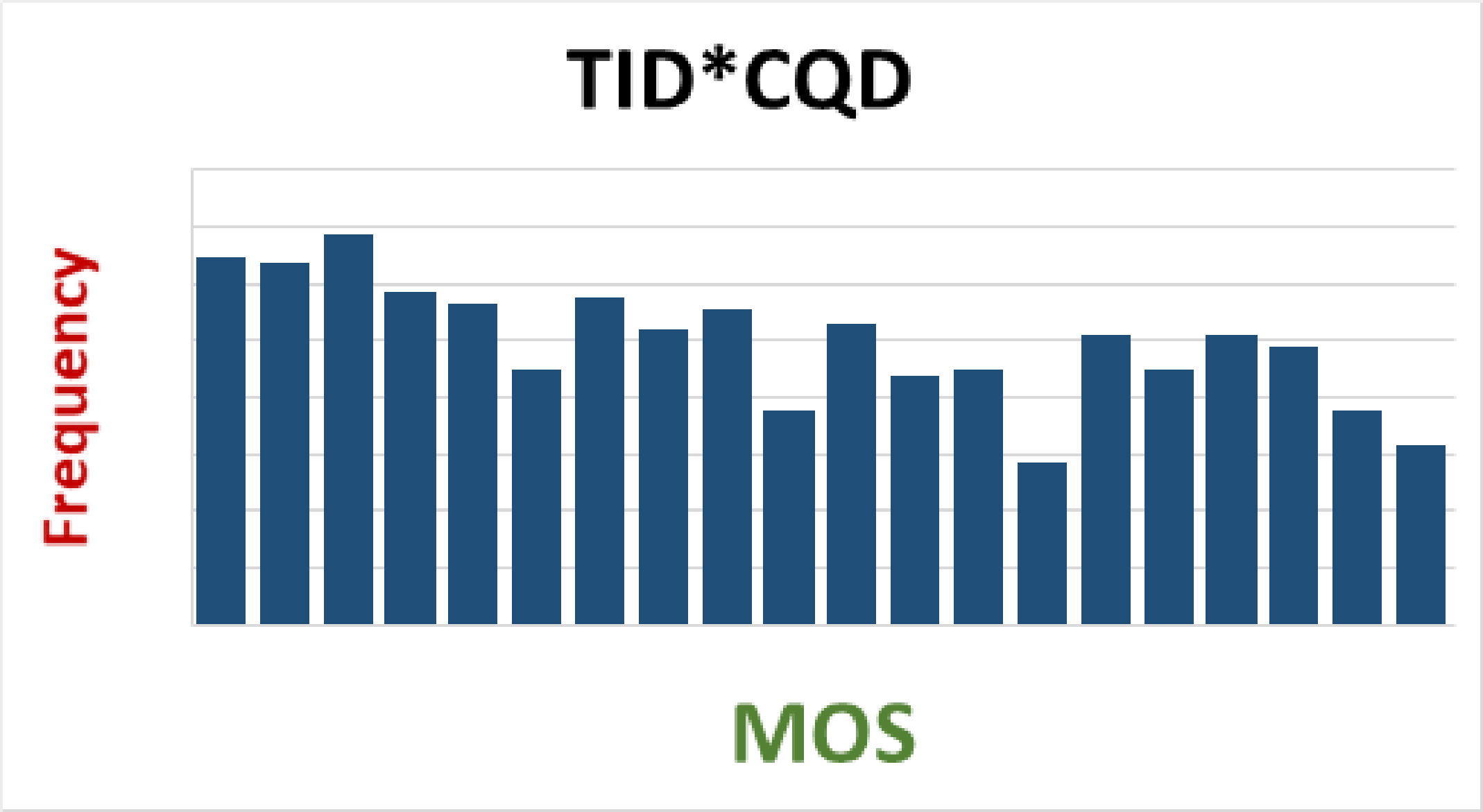}
  \includegraphics[width=0.23\textwidth]{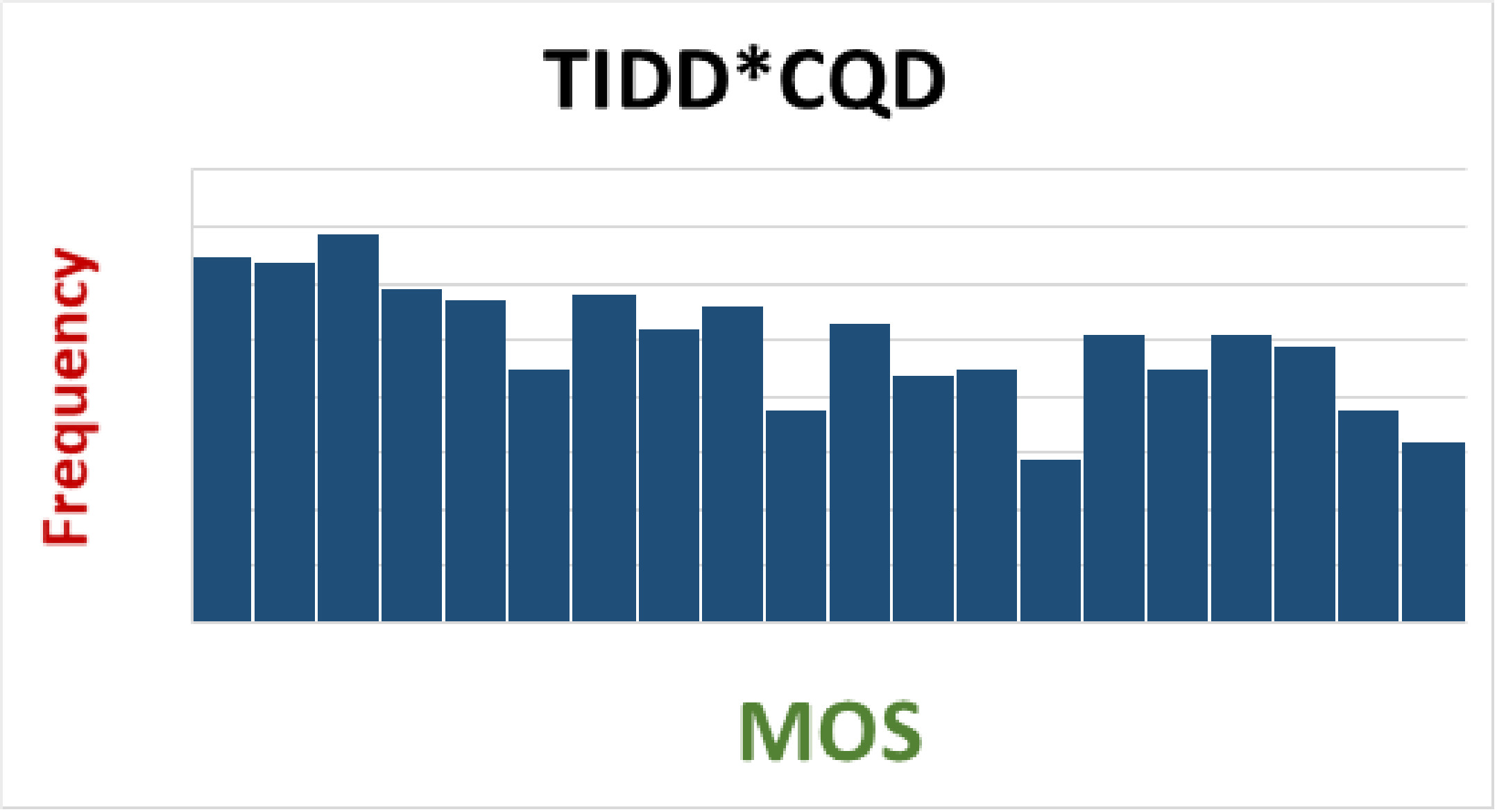}
\caption{Histograms of MOS in the databases taken into account.}
\label{fig:2}       % Give a unique label
\end{center}
\end{figure}

TID2013 contains 25 reference color images (distortion-free, etalon), see Figure\ref{fig:1}, rows 1-5), where 24 source images were obtained (by cropping) from Kodak database \cite{85} and the 25th reference image was artificially created and added in such a way to obtain different content. The fixed size of all images 512x384 pixels is selected as the adequate for the simultaneous displaying of two or three images at the computer monitor screen. 
The distorted images are generated by 24 types of distortions: \#1 - additive Gaussian noise, \#2 - additive noise in color components, \#3 - spatially correlated noise, \#4 - masked noise, \#5 - high-frequency noise, \#6 - impulse noise, \#7 - quantization noise, \#8 - Gaussian blur, \#9 - image denoising, \#10 - JPEG compression, \#11 - JPEG2000 compression, \#12 - JPEG transmission errors, \#13 - JPEG2000 transmission errors, \#14 - non-eccentricity pattern noise, \#15 - local block-wise distortions of different intensity, \#16 - mean shift (intensity shift), \#17 - contrast change, \#18 - change of color saturation, \#19 - multiplicative Gaussian noise, \#20 - comfort noise, \#21 - lossy compression of noisy images, \#22 - image color quantization with dither, \#23 - chromatic aberrations, \#24 - sparse sampling and reconstruction.
Fifth distortion levels for all test images and all distortion types are present for a total of 3000 distorted images, i. e. 25 test images with 24 types and 5 distortion levels in opposite to 1700 distorted images in TID2008 \cite{25} with 25 reference images with 17 types of distortions and 4 distortion levels. Since TID2008 is a subpart of TID2013 and the methodology to obtain MOS has been improved in TID2013, TID2008 will be disregarded in this paper. In the following, for short, we refer at TID2013, simply by TID.\\
All images are saved in the database in a bitmap format without any compression. File names are organized in such a manner that they indicate a number of the reference image, then several distortion's types, and, finally, some distortion's level: "iXX\_YY\_Z.bmp". In TID the two types of distortion related to CQ are quantization noise (\#7) and image color quantization with dither (\#22). The distorted images are obtained by Matlab functions. The classical uniform quantization algorithm is employed to obtain the color quantization (\#7) with a variable number of colors (27, 39, 55 and 76 colors) while CQ with dither (\#22) is modeled using the Matlab function rgb2ind which converts the RGB image to the indexed image using dither. The number of quantization levels is selected individually for each test in the interval [2, 380].\\
TID provides subjective scores, in terms of MOS, for comparing the performance between fidelity measures. The MOS values from TID is collected using a methodology known in psychophysics as Two Alternative Forced Choice (2AFC) match to sample. In 2AFC three images are displayed (the reference and two distorted images) and an observer selects one of the two distorted images which they judge as more similar to the reference, i.e. human observers are asked to select between two images the image that perceptually differs less from a reference image \cite{86}. Thus, the evaluation is made in terms of the presented current stimuli. Since the 2AFC was made within the selected subset of the TID, the MOS scores designated to that subset are a measure of the color difference to the reference image perceived by the observers. Therefore, TID allows the individual analysis of certain distortion type or a subset of distortion types. The subject judgment is expressed with 5 gradations: “Bad”, “Poor”, “Fair”, “Good” and “Excellent” \cite{26}.\\
The validity of the subjective test results was verified by a screening of the results performed according to Annex 2 of ITU-R Rec. BT.500 \cite{34} that defines how to carry out the subjective quality test. Experiment participants, mostly students and in minor extended tutors and researchers, were instructed before starting experiments. The subjective tests via the Internet have been done in different conditions using different monitors both LCD and CRT, mainly 19 inches and more with the resolution 1152x864 pixel. The used distance from monitors is the best comfort for them. Then, the obtained results are averaged for each reference image. Thus, the obtained MOS has to vary from 0 to 9 and its larger values correspond to better visual quality, although there are no MOS values larger than 7,5, see Figure \ref{fig:2}. 
As confirmed also in \cite{23}, the averaged deviation standard of TID is sufficiently small to guarantee that most subjects have similar ratings about image quality and, consequently, the subjective evaluation yields a highly reliable score. Moreover, in TID twenty-eight different commonly used metrics are taken into account for the evaluation and calculated using Metrix MUX Visual Quality Assessment package \cite{87} (see also \cite{26} for more details).
\subsection{Color Quantization Database (CQD)}
\label{sec3.2}
\begin{figure}
\begin{center}
\includegraphics[width=0.23\textwidth]{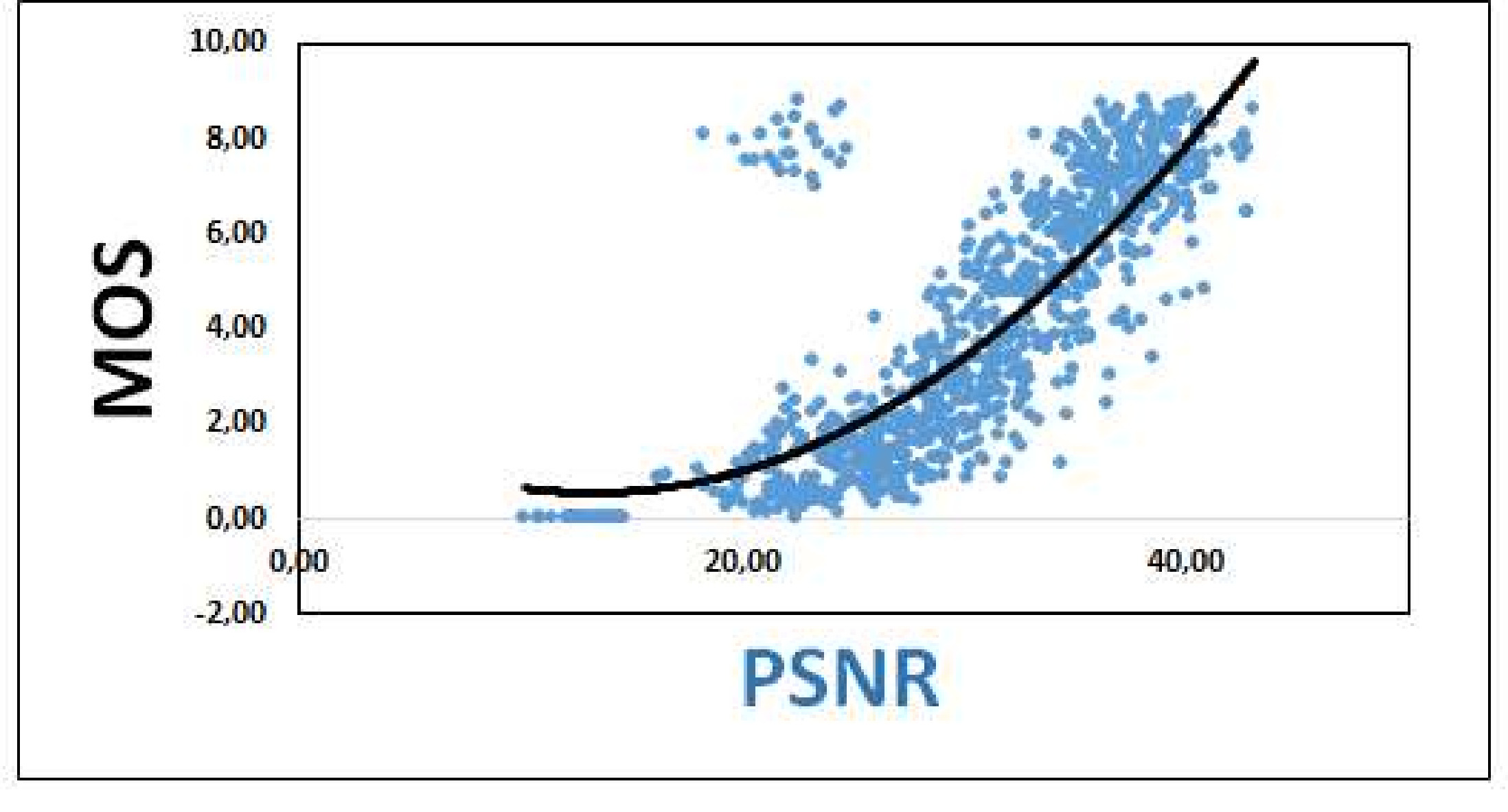}
  \includegraphics[width=0.23\textwidth]{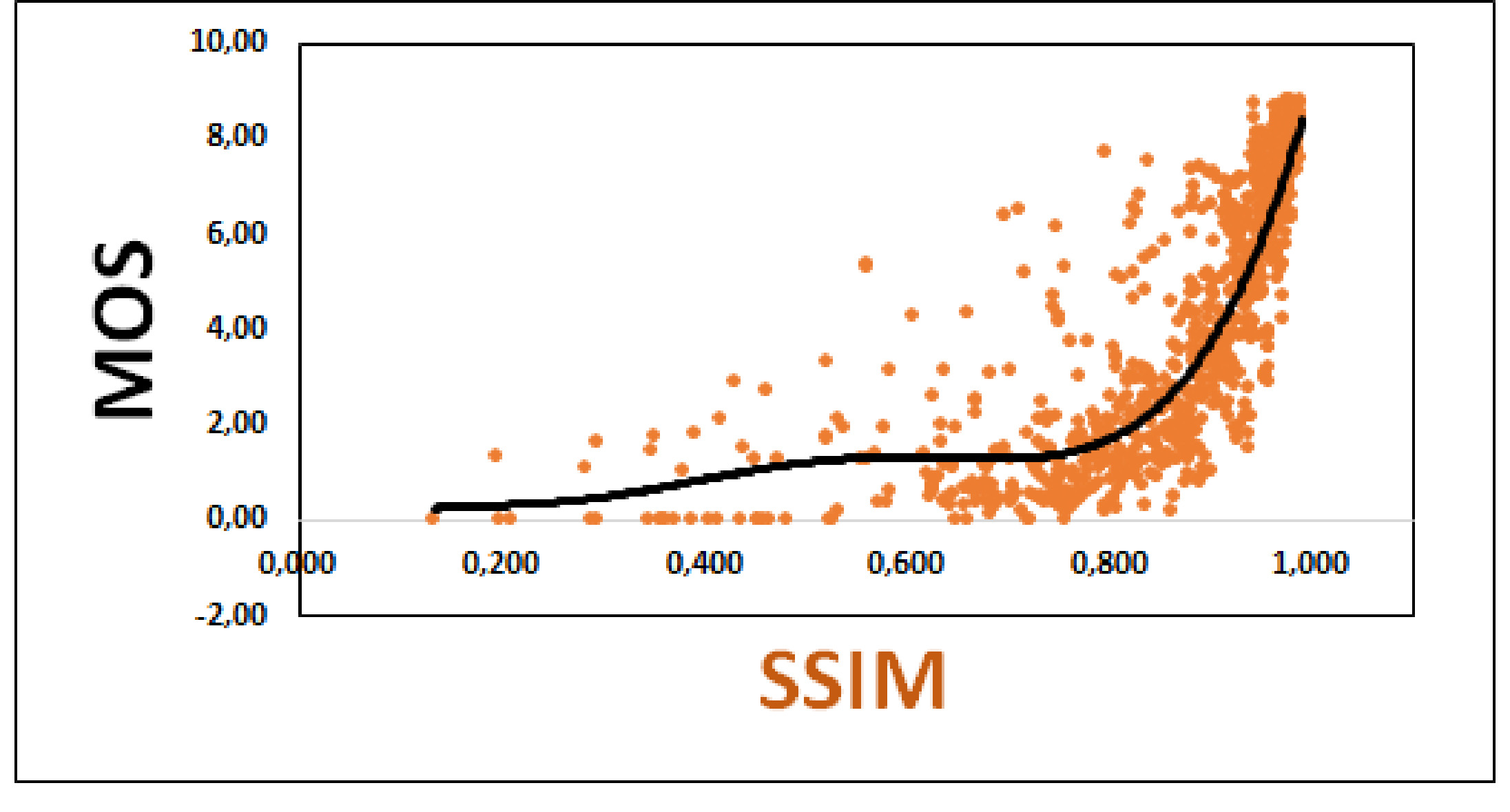}
 \includegraphics[width=0.23\textwidth]{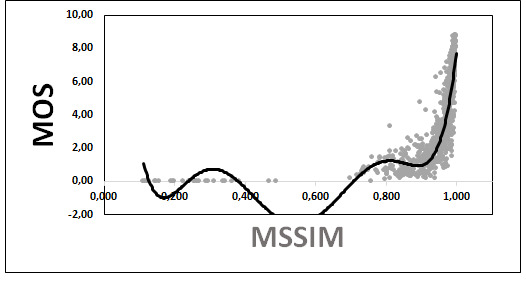}
  \includegraphics[width=0.23\textwidth]{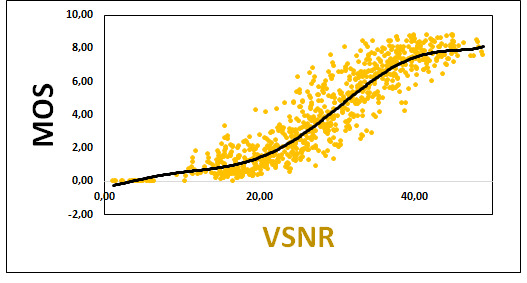}
 \includegraphics[width=0.23\textwidth]{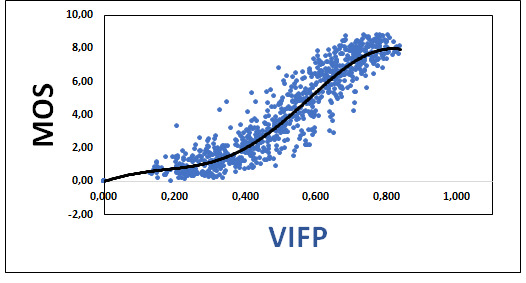}
  \includegraphics[width=0.23\textwidth]{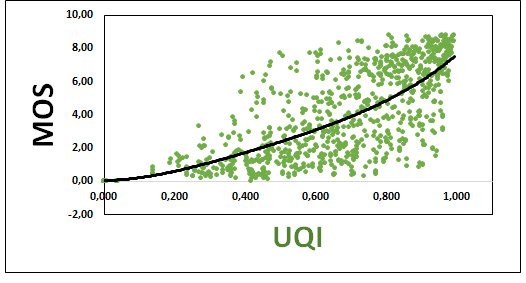}
 \includegraphics[width=0.23\textwidth]{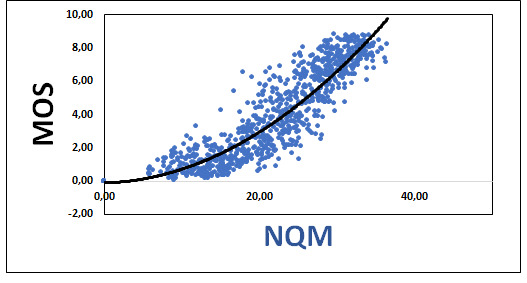}
 \includegraphics[width=0.23\textwidth]{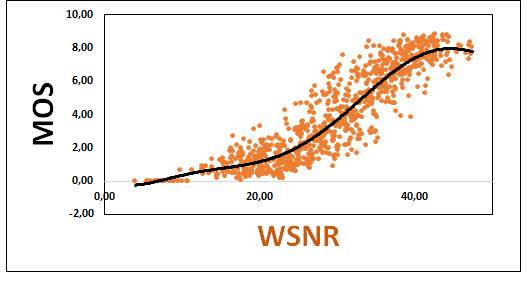}
  \includegraphics[width=0.23\textwidth]{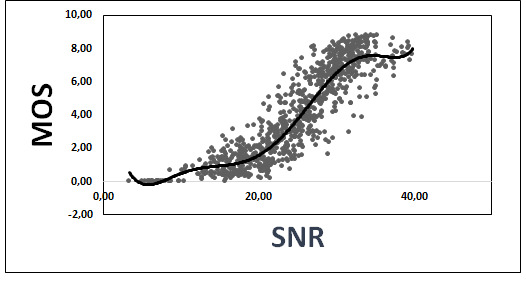}
\caption{Scatter plots of quality measures for database CQD.}
\label{fig:3}       % Give a unique label
\end{center}
\end{figure}

CQD consists of 25 reference images (see Figure 1, rows 6-10) collected from the Web, based on the number of segments and the number of colors to reflect a variety of image contents as an important object, uniform regions, slowly varying color gradients, edges, and high level of details. The fixed size of all images is 512x512 pixels. The unique considered distortion is CQ. For each reference image, seven levels of quantization are generated: 4, 8, 16, 32, 64, 128, 256 colors by five popular CQ methods, for a total of 875 distorted images. The popular CQ methods are: Kmeans method \cite{41}, Median Cut method \cite{42}, Octree \cite{43}, Wu’s method \cite{44}, Dekker’s SOM \cite{45}.\\
All images are saved in the database in png format without any compression. The reference images are numbered and the file names of the distorted image are organized in a way that they indicate some colors X in the quantized image and the number of the reference image Y. For example, the name 32colors\_7.png means the 7th reference image quantized to 32 colors.\\
CQD provides subjective scores, in terms of MOS, for comparing the performance between fidelity measures. To evaluate the quality of the quantized images, a subjective quality test is used in which many human subjects are asked to judge the quality of the sequence images. The subjective tests are based on the recommendations given by the ITU-R Rec. BT.500 \cite{34}. Preliminarily, a group of twenty students was instructed concerning what they are going to see, what they have to evaluate, and how they have to express their opinion. The volunteers also have been shown some examples of quantized images, different from those used in the actual experiment,  for various quantization levels. During the testing phase, each subject is watching two images (reference and test) at the same time and is asked to assess the quality of the test image to the reference image by simply dragging a slider on a quality scale with range [0,100] labeled and divided into five equal categories: “Bad”, “Poor”, “Fair”, “Good”, and “Excellent”. The psychometric experiments were conducted in an environment with normal indoor illumination and the display monitors were all 19 inch CRT, with all approximately the same age and with the same display settings. To compare with the obtained MOS of TID, the normalized values, according to the equation (7) in the successive section 4.3, are shown in Figure 2 with a larger value equal to 8,78. Also for CQD, the averaged deviation standard is sufficiently small to guarantee that most subjects have similar ratings about image quality and consequently the subjective evaluation yields a highly reliable score.\\
Several well-known objective image quality metrics are considered in \cite{27} where the publicly available implementations of these quality metrics are used. In Figure \ref{fig:3}, scatter plots for the subjective MOS versus the quality score from different image quality metrics are shown. For some evaluations, in the following, we refer also to some subparts of CQD, namely CQD-Median, CQD-Kmeans, CQD-Octree, CQD-Wu, CQD-Som, relative to data of CQD obtained by the corresponding already mentioned CQ methods.\\
In summary, the main common and uncommon features of TID and CQD are the following. Differently from TID, in CQD the number of colors is always the same for each distortion level (multiple by 2). TID and CQD have both complied with ITU recommendations and have 5 variability classes, although in different intervals: [1,9] for TID and [0,100] for CQD. The employed metrics have a 9/10 ratio.
% For one-column wide figures use (NOTA fig4 e fig 5 originali sono invertite)
\section{Evaluation Method}
\label{Evaluation Method}
\subsection{Motivations and objectives}
\label{sec4.1}
IQA has become an active area in color image processing due to its wide range of applications such as color correction \cite{88}, color quantization \cite{7}, color mapping \cite{89}, color image similarity, and retrieval \cite{90}. In particular, in color quantization, adequate color difference assessment can be used to find the appropriate quantization step size and/or range of displayable colors to obtain the image reproduction with the minimum loss in perceived colors. On the other hand, as previously mentioned, while many quality measures for natural scene color images have been proposed, there has not yet been any rigorous and satisfying investigation into the performance comparison of the existing measures \cite{35,91,92,93,94} and little studied in the literature turns out to be the performance evaluation of these measures in the presence of CQ corruption. Moreover, as mentioned in section 3, these investigations are tested on few public databases which contain multiple distortions, mostly related to spatial distortions, often combined with the distortion related to the color, and the available data are not adequately structured to carry out such CQ-related assessment. An adequate assessment of existing metrics in the presence of distortions due merely to the CQ is therefore missing.\\
For the above considerations, the main goals of this paper are a) to determine the appropriate available databases; b) to attempt to fill this gap by assessing specific full reference metrics in the presence of CQ distortion; c) to propose a novel performance evaluation methodology based on recommendations for best practices.
\subsection{Databases integration: pros and cons}
\label{sec4.2}
Predictive performance evaluation tools of different full-reference image quality metrics are significantly impacted by the choice of the image quality database \cite{84}. Hence, the first step to evaluate the existing quality measures for CQ is to determine which databases are convenient to perform such an assessment. As aforementioned, CQD excluded, there is a lack of a suitable database which addresses specifically CQ alterations (building such a database is outside the scope of this paper), then to achieve our goals, we have to use the publicly available databases by isolating data with CQ degradation.\\
Therefore, the initial step is to consider only publicly available databases containing images that undergo a CQ process. Thus, we have considered TID and we have extracted images with CQ and with CQ followed by dithering from TID. In the following, these databases are indicated by TID* and TIDD*, respectively.\\
Note that the evaluation of quality measures on separate databases is in principle explanatory. Usually, many authors train their quality measure methods on a database typically 
maximizing individual correlations or minimizing prediction errors. In that case, each database has a high dimensionality and is characterized by many significant features such as do not require the use of data extension through the combination (or fusion) of multiple databases. Some authors prefer to merge databases only when it is possible to statistically prove that the data are extracted from the same population \cite{95} since an eventual positive answer to this theoretical question does not require further prior investigations and ensures that a merge is possible and has its statistical reliability. Conversely, a negative response requires a evaluation performed on the database obtained by fusion that involves estimating the quality measures on each database and the integrated databases. This assessment can validate the fusion if the obtained results on both the individual and the integrated databases are compatible with each other.\\
In our case, to see if the data from both databases (CQD and TID) can be considered as coming from the same population, on the MOS of each database, the Kolmogorov- Smirnov test \cite{96} and Manh-Whitney test \cite{97} have been performed. The result has confirmed that the data from both databases cannot be a priori considered as coming from the same population and therefore, in principle, not directly mergeable. However, the number of images of each database is small and instances of CQ distortions are quite limited, i.e. the available databases have a reduced cardinality and are little representative of the CQ phenomenon. In particular, the images with quantization noise are obtained by applying a simple CQ method with quantization levels varying within a prefixed range (TID) or by applying different well-known methods with a prefixed number of quantization levels: 4, 8, 16, 32, 64, 128, 256 (CQD).\\
In our opinion, the possible fusion of these databases could describe the CQ phenomenon through a more representative database and to obtain a more precise and an adequate assessment could result  easier since it relates to different instances of the same phenomenon obtained under different conditions and through different methods. In support of this argument, analysis of image quality in slightly varying conditions provide reasonably good verification of quality metrics since non-identical conditions of experiments take into account how visual quality is assessed in a priori unknown conditions in everyday practice, as also stated in \cite{26}. \\
On the other hand, given the reduced cardinality and non-heterogeneity of the data in each database, the assessment carried out separately on each database is rather not significant and it would be desirable to be able to evaluate the metrics on a larger heterogeneous database, to verify if the result is compatible with it. Therefore, in the absence of a more appropriate database, we feel is reasonable to evaluate the various metrics on a database obtained through the integration of the available data.\\
A not negligible fact is that the image sets of the TID and CQD, and then TID subparts and CQD, do not have common images. As a result, if one part it is not possible to make an exact comparison on a specific image in the respective databases, on the other it favors their eventual integration because their fusion does not lead to image duplication and allows actually to increase the cardinality of the test image set.\\
For this integration, it also plays the fact that the authors who proposed the TID and CQD databases claim to have complied with the ITU \cite{34} guidelines and therefore, in principle, such databases (and TID subparts) are comparable at the experimental level. Of course, compliance with ITU positions is a necessary but not sufficient condition to ensure complete data reliability.\\
Another positive factor is that CQD and TID (subparts included) have 9 out of 10 metrics in common and have 5 variability classes, although variability classes are in different ranges. This allows us to compare the values of the metrics obtained after an appropriate normalization rule (see section 4.3 for more details).\\
In addition to the increase in cardinality and the heterogeneity that produces a better representation of the CQ phenomenon closer to real sample, the use of such an integrated database offers the advantage of widening the sample of the type of humans who have participated in the experiment although the experimental conditions in which the MOS have been computed are only "partially" equal. As a consequence, the database integration, while on the one hand leads to an increase in instability due to experimental conditions, on the other leads to an increase in the data likelihood due to a different sample of the humans involved and therefore an increase of the reliability of the MOS values and makes the performance evaluation statistically more significant. Of course, this factor should be fully taken into account in the evaluation process.\\
Due to the aforementioned reasons, different databases are considered to evaluate the quality measures for CQ degradation.
\subsection{Methodology}
\label{sec4.3}
The two aforementioned databases (see section 3) have been prepared as follows. CQD has been fully extracted and its data has been re-arranged so that it can be compared with TID.\\
Since the considered metrics values were used for CQD in \cite{27} but are not publicly available, these values have been recalculated using, as done in \cite{26}, the Matlab code Metrix MUX Visual Quality Assessment package \cite{87} so to favorite the comparison with TID data.\\
From TID, data have been extracted to form two sub-databases by selecting those relative to the color quantization distortion (\#7) and color quantization followed by dithering (\#22):\\
- TID*: containing TID images with CQ distortion (\#7) and the corresponding MOS;\\
- TIDD*: containing TID images with CQ distortion plus dithering (\#22) and the corresponding MOS.\\
The availability of these new databases has allowed the assessment not only regarding CQ degradation but also CQ degradation with dithering. The remaining 21 distortion types contained in TID are not used since not even those affecting color because they incorporate also spatial distortions which typically impact the quality of the image much more strongly than color alteration. Therefore, the human scores would be then more likely predominantly influenced by the spatial distortions and not the color ones.\\
Although all considered databases and their subparts have 5 classes of variability, the MOS values in TID, TID*, TIDD* varies in the range [0,9] while those in CQD varies in the range [0,100].\\
To compare the data and to be ready for the fusion, it has been therefore necessary to normalize the MOS of CQD, i. e. to make a data transformation such that data refer to a common field of variation. \\
Therefore, let  $X_{min}$ and $X_{max}$ be respectively the minimum and the maximum value of the variation interval (in this case $[0,100]$) and let $K$ be the maximum value of the new interval, each value $N_i$ is replaced by the value given by the following formula:
\begin{equation}
N_i= \displaystyle  \frac{ K(X_i-X_{min})}{ (X_{max}- X_{min})}
\end{equation}
Note that by this normalization formula, the values range in the interval $[0, K]$. In our case, $K=5$ and each value $N_i$  of $CQD$ is multiplied for $9/100$. After performing all these preliminary steps, the integration of the databases has been done by obtaining TID*CQD and  TDD*CQD which include the corresponding MOS and quality measures. This integration has allowed performing the evaluation for CQ degradation and for CQ degradation with dithering on data sets characterized by greater cardinality and heterogeneity. This property characterizing the integrated datasets, as mentioned above, is adequate for the performance of the evaluation process.
\subsection{Features of employed databases}
\label{sec4.4}
\begin{figure}
\begin{center}
\includegraphics[width=0.23\textwidth]{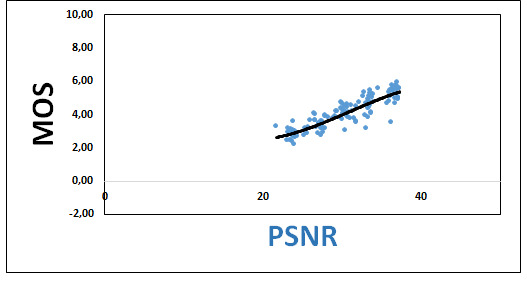}
  \includegraphics[width=0.23\textwidth]{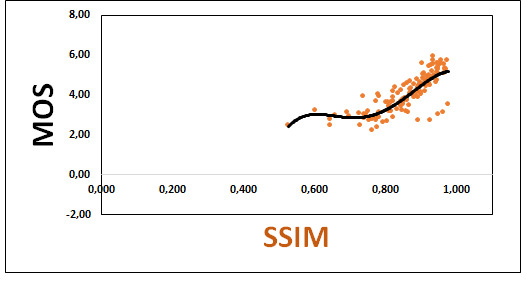}
 \includegraphics[width=0.23\textwidth]{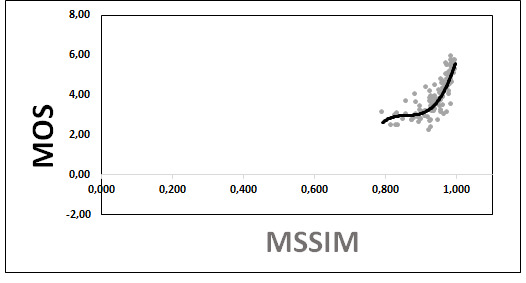}
  \includegraphics[width=0.23\textwidth]{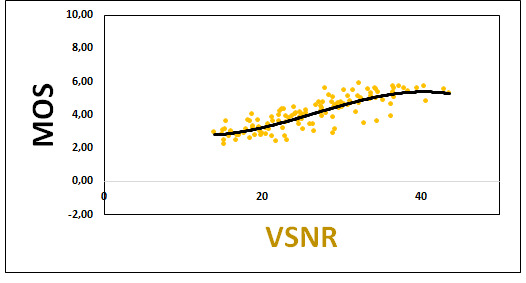}
 \includegraphics[width=0.23\textwidth]{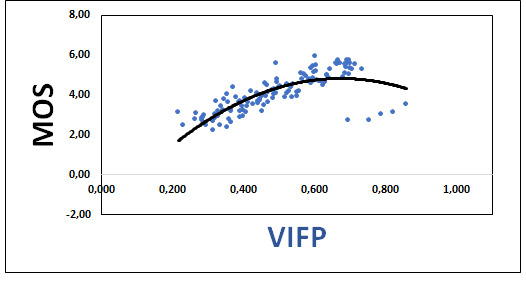}
  \includegraphics[width=0.23\textwidth]{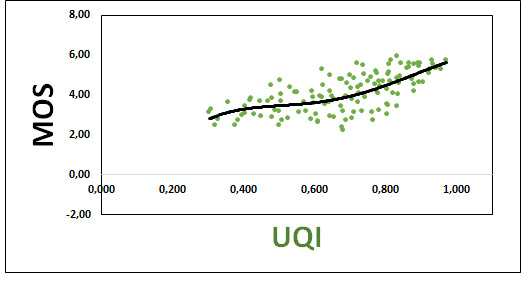}
 \includegraphics[width=0.23\textwidth]{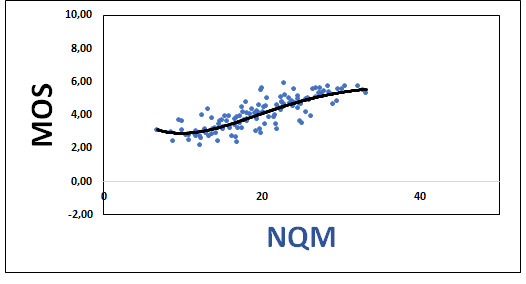}
 \includegraphics[width=0.23\textwidth]{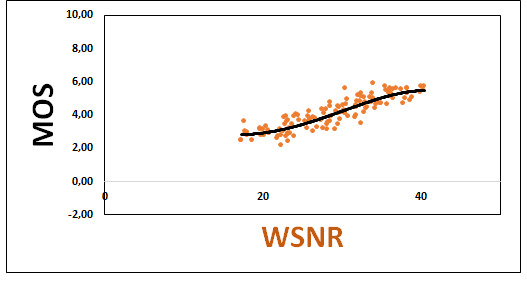}
  \includegraphics[width=0.23\textwidth]{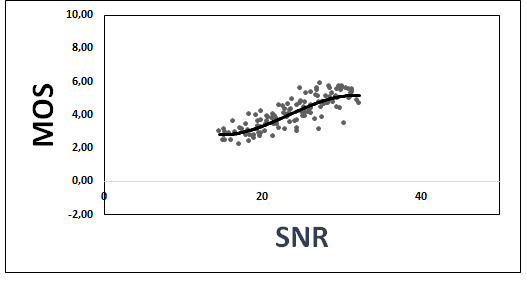}
\caption{Scatter plots of quality measures for database TID*.}
\label{fig:4}       % Give a unique label
\end{center}
\end{figure}

\begin{figure}
\begin{center}
\includegraphics[width=0.23\textwidth]{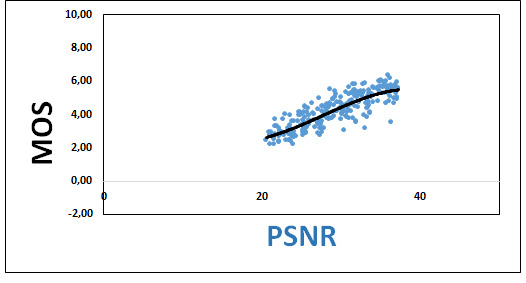}
  \includegraphics[width=0.23\textwidth]{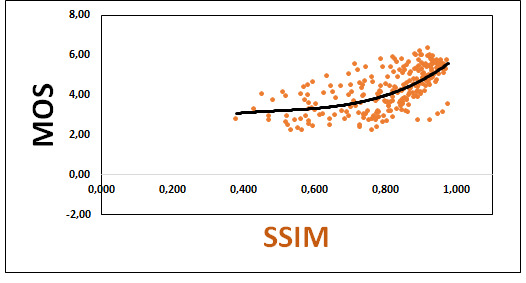}
 \includegraphics[width=0.23\textwidth]{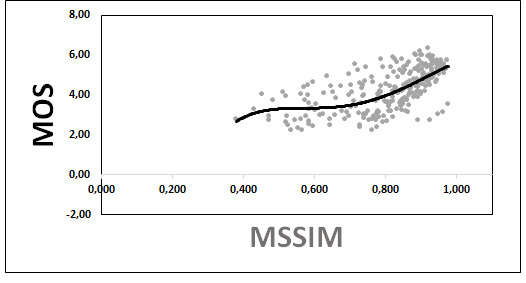}
  \includegraphics[width=0.23\textwidth]{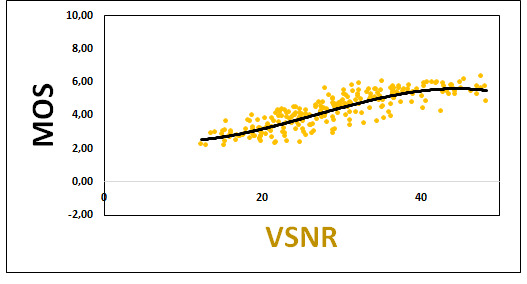}
 \includegraphics[width=0.23\textwidth]{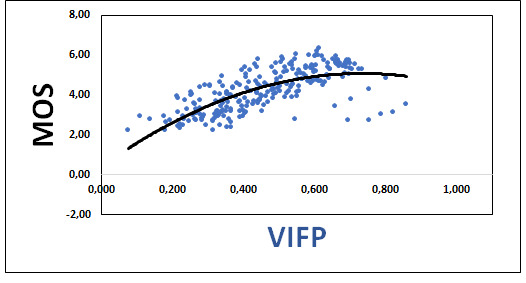}
  \includegraphics[width=0.23\textwidth]{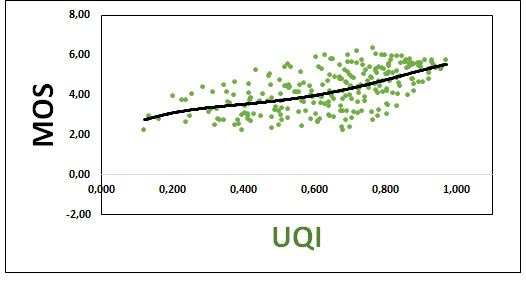}
 \includegraphics[width=0.23\textwidth]{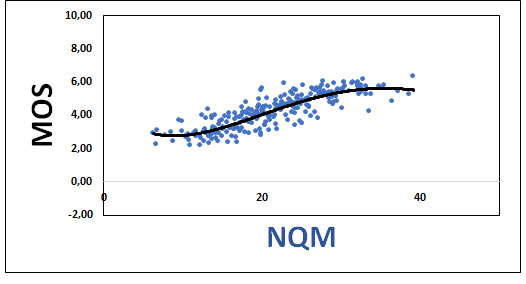}
 \includegraphics[width=0.23\textwidth]{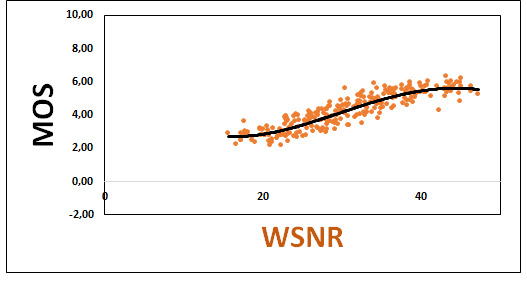}
  \includegraphics[width=0.23\textwidth]{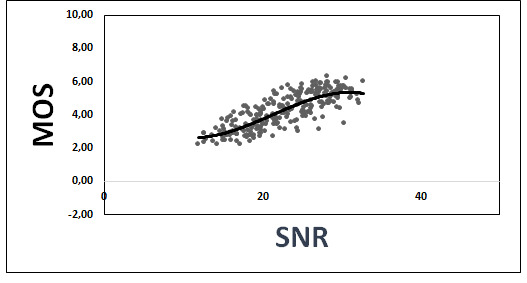}
\caption{Scatter plots of quality measures for database TIDD*.}
\label{fig:5}       % Give a unique label
\end{center}
\end{figure}

\begin{figure}
\begin{center}
\includegraphics[width=0.23\textwidth]{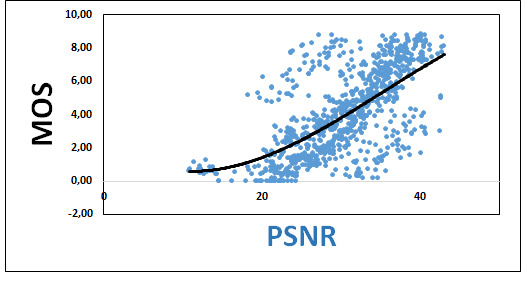}
  \includegraphics[width=0.23\textwidth]{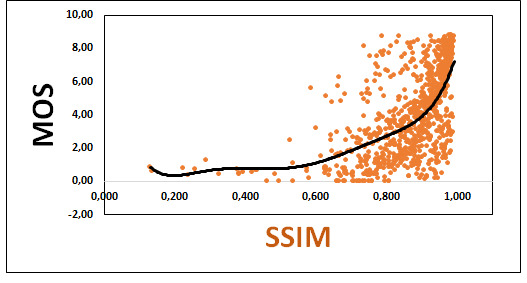}
 \includegraphics[width=0.23\textwidth]{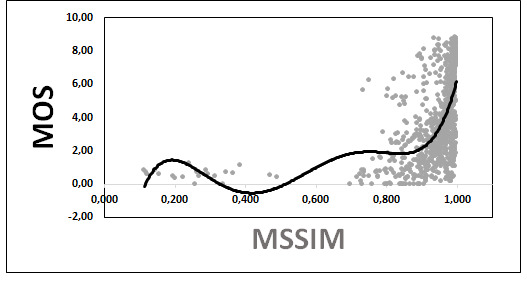}
  \includegraphics[width=0.23\textwidth]{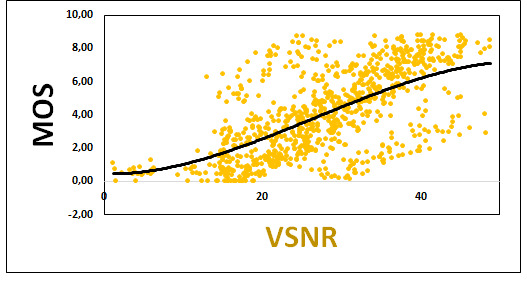}
 \includegraphics[width=0.23\textwidth]{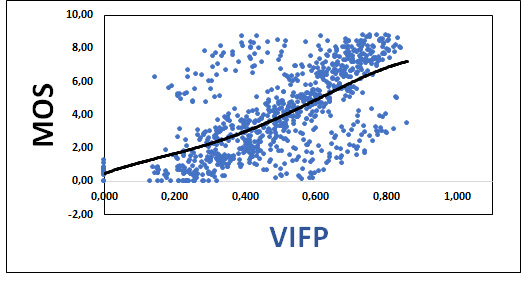}
  \includegraphics[width=0.23\textwidth]{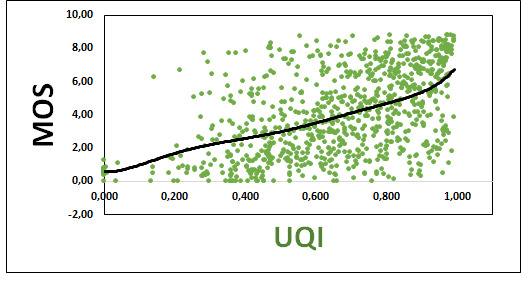}
 \includegraphics[width=0.23\textwidth]{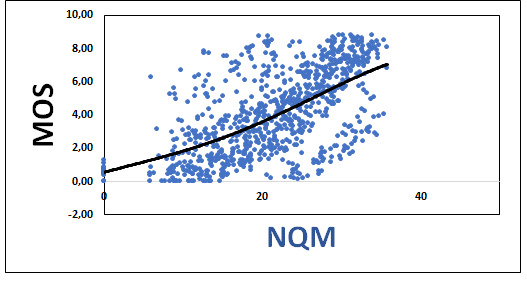}
 \includegraphics[width=0.23\textwidth]{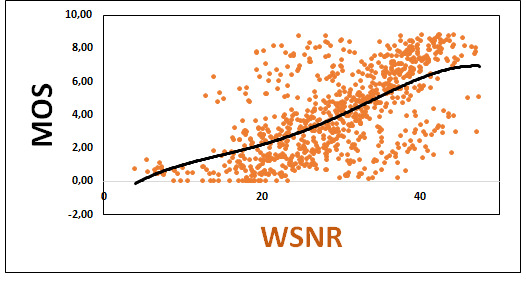}
  \includegraphics[width=0.23\textwidth]{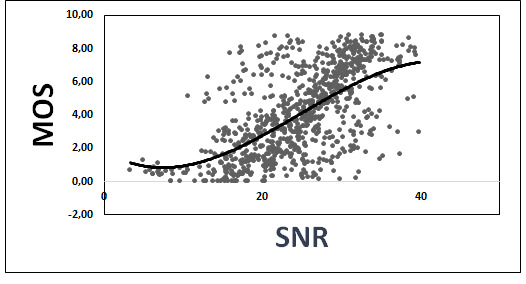}
\caption{Scatter plots of quality measures for database TID*CQD.}
\label{fig:6}       % Give a unique label
\end{center}
\end{figure}

\begin{figure}
\begin{center}
\includegraphics[width=0.23\textwidth]{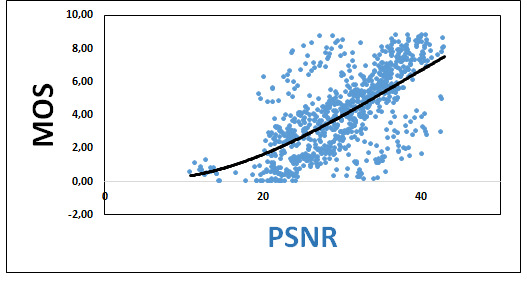}
  \includegraphics[width=0.23\textwidth]{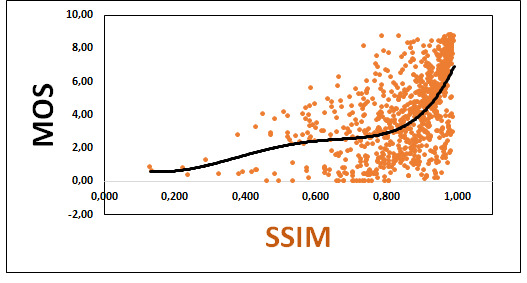}
 \includegraphics[width=0.23\textwidth]{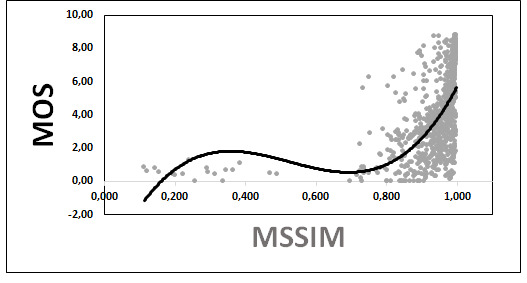}
  \includegraphics[width=0.23\textwidth]{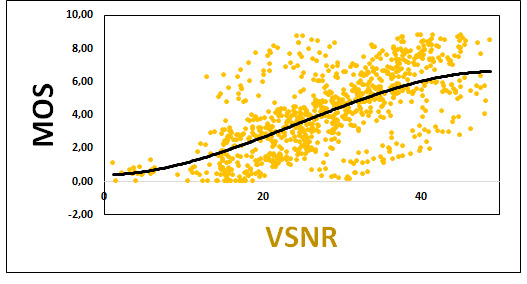}
 \includegraphics[width=0.23\textwidth]{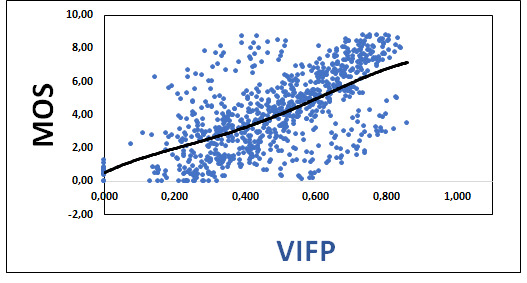}
  \includegraphics[width=0.23\textwidth]{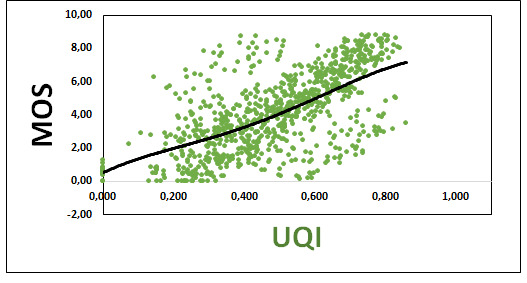}
 \includegraphics[width=0.23\textwidth]{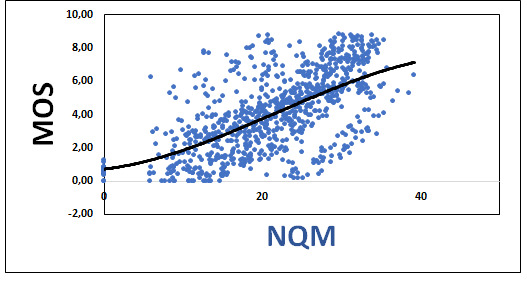}
 \includegraphics[width=0.23\textwidth]{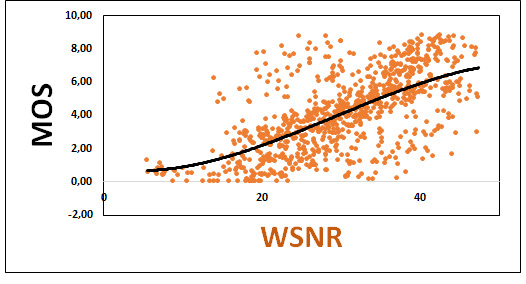}
  \includegraphics[width=0.23\textwidth]{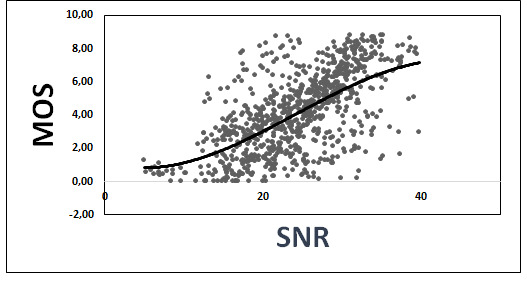}
\caption{Scatter plots of quality measures for database  TIDD*CQD.}
\label{fig:7}       % Give a unique label
\end{center}
\end{figure}

TID* contains the 25 reference images of TID of fixed size 512 x 384 pixels plus the distorted images generated by quantization noise (\#7) with fifth distortion levels for a total number of 125 images. Being TID* a subpart of TID, of course, the images belonging to TID* have the same features of TID in terms of format, file names, color quantization noise. Moreover, the number of quantization levels is selected individually for each test in the interval [2, 380], while the corresponding MOS are extracted from TID. The nine selected metrics relative to the new dataset TID* are extracted from TID. In Figure 2 and Figure \ref{fig:4}, the histogram of MOS with a maximum value equal to 5.91 and scatter plots for the subjective MOS versus the quality score for TID* are shown, respectively. 
TIDD* contains the 25 reference image of TID of fixed size 512 x 384 pixels, the distorted images generated by quantization noise (\#7), and color quantization with dithering (\#22) with fifth distortion levels for a total number of 250 images. It has the same features of TID* in terms of format, file names, color quantization noise, and the number of quantization levels is selected individually for each test in the interval [2, 380], while the corresponding MOS are extracted from TID. Even for TIDD* the corresponding MOS are extracted from TID and the nine selected metrics relative to the new dataset TIDD* are extracted from TID. In Figure 2 and Figure \ref{fig:5}, the histogram of MOS with a maximum value equal to 6.33 and Scatter plots for the subjective MOS versus the quality score for TIDD* are shown, respectively. 
TID*CQD contains the 25 reference image of TID of the fixed size 512 x 384 pixel, the distorted images generated by quantization noise (\#7) at fifth distortion levels for a partial total of 125 images plus the 25 reference images of CQD of a fixed size of 512 x 512 pixels, the distorted images generated by the five popular CQ methods. at seventh levels of quantization (4, 8, 16, 32, 64, 128, 256 colors) for a partial total of 875 distorted images, and hence with a total of 1000 distorted images. In TID*CQD the images have different file names, the number of quantization levels and format type (.bmp or .png) depending on the database from which they have been derived, while the corresponding MOS are respectively extracted from corresponding database TID* and the normalized values of CQD according to the equation (8). In Figure 2 and Figure \ref{fig:6}, the histogram of MOS with a maximum value equal to 8.79 and scatter plots for the subjective MOS versus the quality score for TID*CQD are shown, respectively. 
Similarly,  TDD*CQD contains the 250 images of TIDD* of fixed size 512 x 384 pixel plus CQD images, of the fixed size of 512 x 512 pixels, the distorted images generated by the five popular CQ methods. at seventh levels of quantization (4, 8, 16, 32, 64, 128, 256 colors) for a partial total of 875 distorted images, and hence with a total of 1125 distorted images. The corresponding MOS are respectively extracted from corresponding database TIDD* and the normalized values of CQD according to the equation (8). In Figure 2 and Figure \ref{fig:7}, the histogram of MOS with a maximum value equal to 8.79 and scatter plots for the subjective MOS versus the quality score for TDD*CQD are shown, respectively. 
Of course, also for TID*CQD and TIDD*CQD, the averaged deviation standard is sufficiently small and comparable with the other databases under consideration to guarantee that most subjects have similar ratings about image quality and consequently the subjective evaluation yields a highly reliable score.\\
From Figure 2, where the histogram of the normalized MOS of all databases are shown, it can be seen that: i) the MOS of CQD, TID*, TIDD*, TID*CQD and  TDD*CQD approximate the uniform distribution most closelyfor those of TID which exhibit an n-shaped distribution; ii) the other database makes the best use of the MOS range without emphasizing any part of the scale since they are characterized by a more uniform distribution. This uniformity of MOS of the considered databases is an important benchmark since a uniform distribution of MOS indicates that the range of rating scales is utilized fully, as reported also in \cite{95}  and \cite{23}. All considered databases are deposited in a publicly available database at -----. The accession numbers will be provided during review and before publication.
\section{Evaluation results}
\label{sec4.5}
\subsection{Evaluation of quality measures}
\label{sec4.5.1}
\begin{figure*}
%\begin{center}
\includegraphics[width=0.5\textwidth]{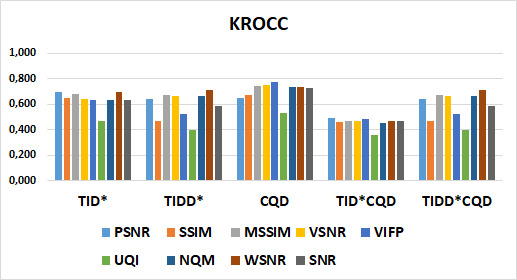}
\includegraphics[width=0.5\textwidth]{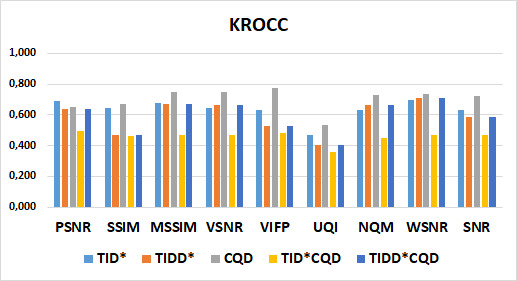}
  \includegraphics[width=0.5\textwidth]{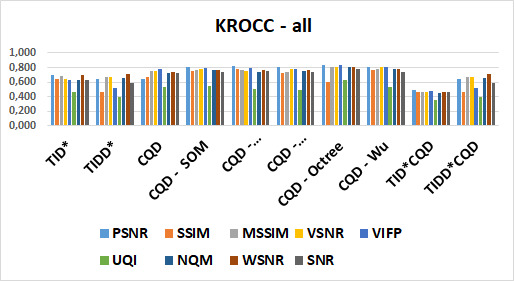}
\includegraphics[width=0.5\textwidth]{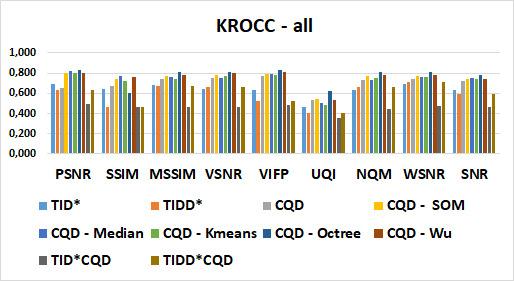}
 \includegraphics[width=0.5\textwidth]{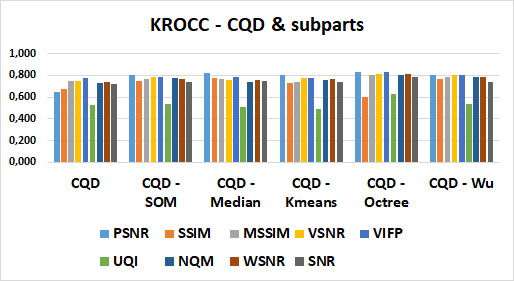}
\includegraphics[width=0.5\textwidth]{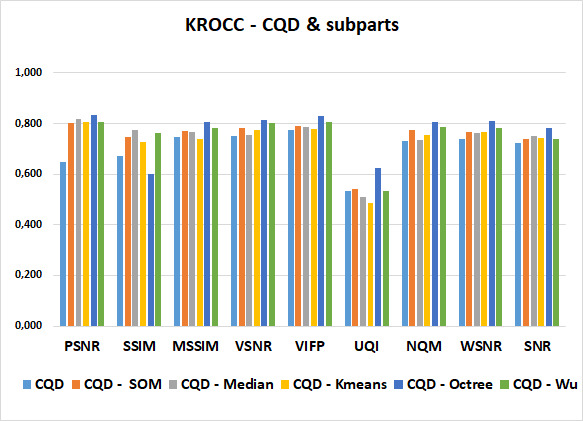}
\caption{Bar charts for KROCC by databases (left) and by quality measures (right).}
\label{fig:8}       % Give a unique label
%\end{center}
\end{figure*}

\begin{figure*}
%\begin{center}
\includegraphics[width=0.5\textwidth]{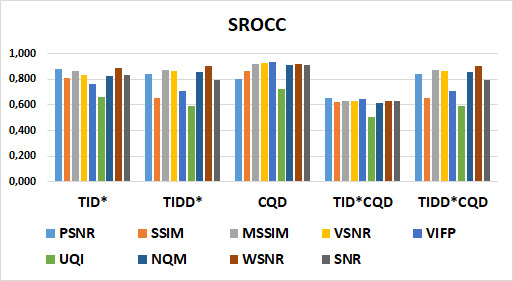}
 \includegraphics[width=0.5\textwidth]{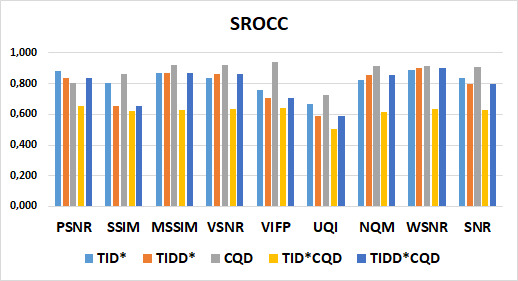}
  \includegraphics[width=0.5\textwidth]{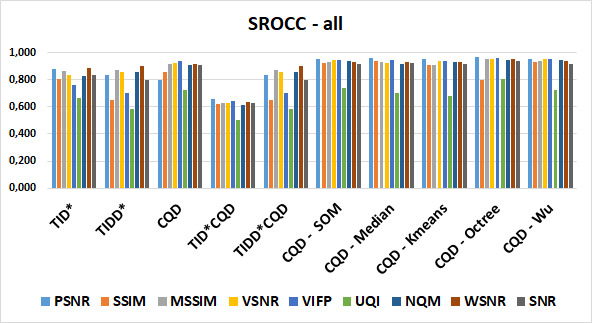}
\includegraphics[width=0.5\textwidth]{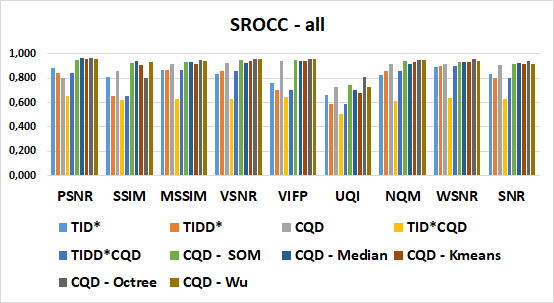}
 \includegraphics[width=0.5\textwidth]{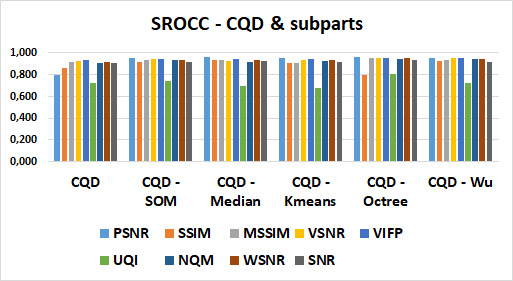}
  \includegraphics[width=0.5\textwidth]{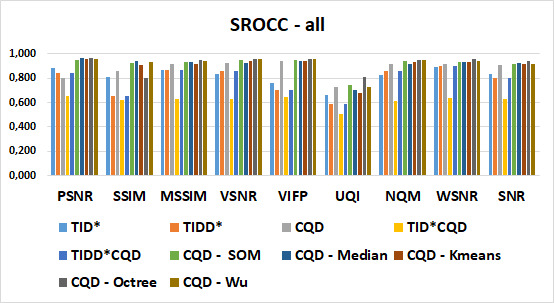}
\caption{Bar charts for SROCC by databases (left) and by quality measures (right).}
\label{fig:9}       % Give a unique label
%\end{center}
\end{figure*}

\begin{figure}
\begin{center}
\includegraphics[width=0.35\textwidth]{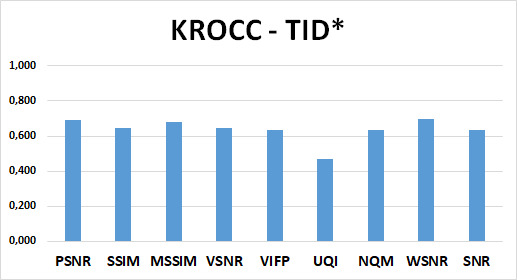}
\includegraphics[width=0.35\textwidth]{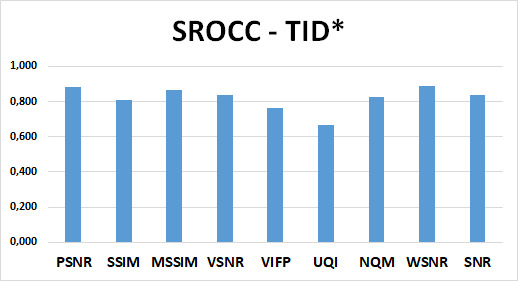}
\includegraphics[width=0.35\textwidth]{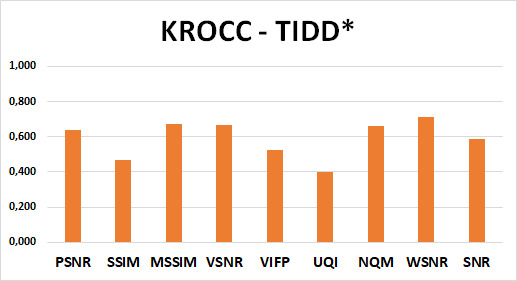}
\includegraphics[width=0.35\textwidth]{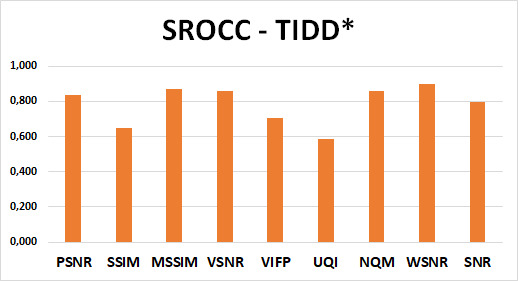}
\includegraphics[width=0.35\textwidth]{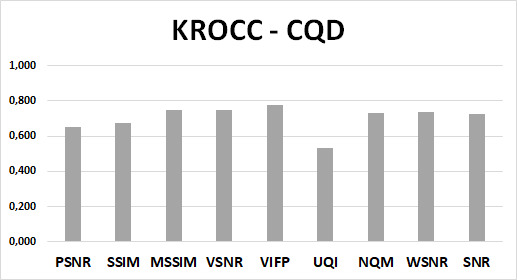}
\includegraphics[width=0.35\textwidth]{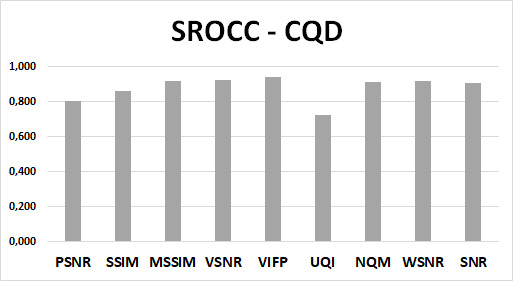}
\includegraphics[width=0.35\textwidth]{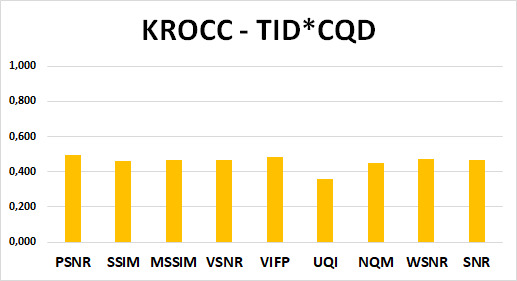}
\includegraphics[width=0.35\textwidth]{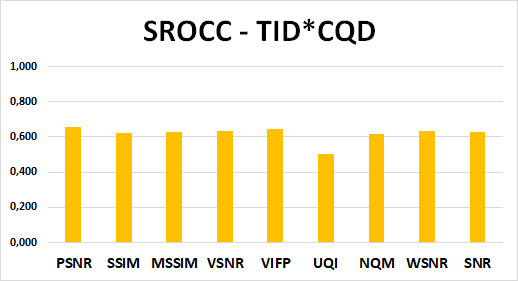}
\includegraphics[width=0.35\textwidth]{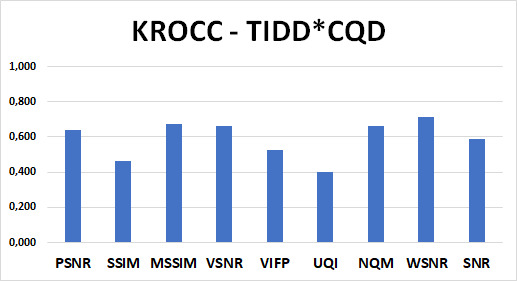}
\includegraphics[width=0.35\textwidth]{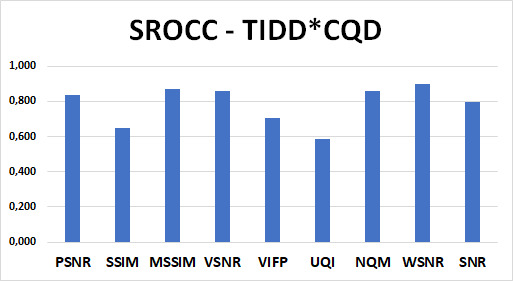}
\caption{Bar charts for KROCC (left) and for SROCC (right) of all quality measures for each database.}
\label{fig:10}       % Give a unique label
\end{center}
\end{figure}

\begin{figure*}
%\begin{center}
\includegraphics[width=0.50\textwidth]{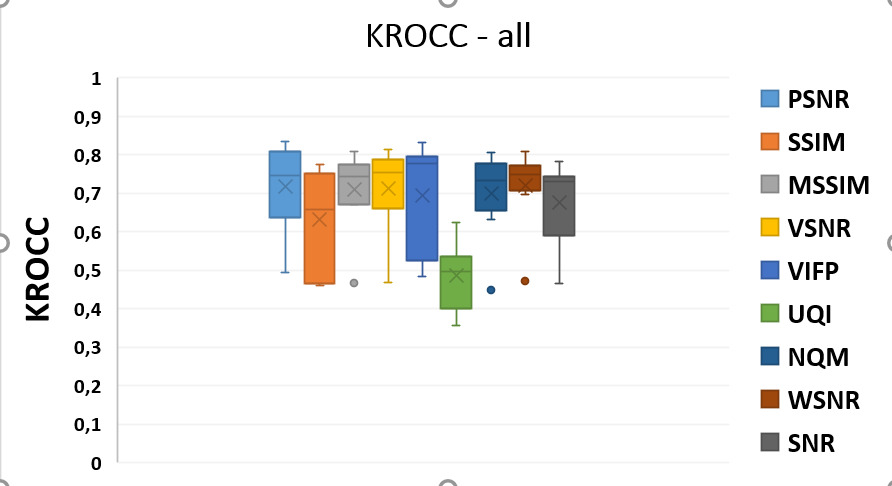}
\includegraphics[width=0.50\textwidth]{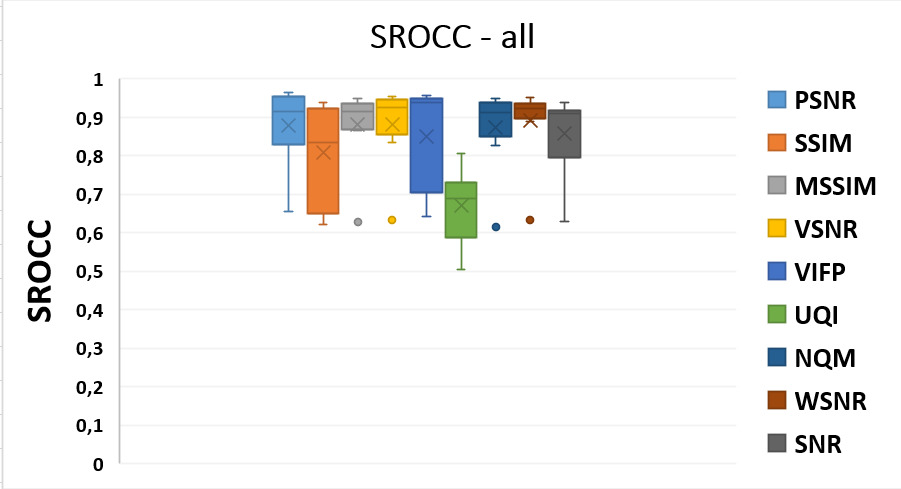}
\caption{Box plot of KROCC for each quality measures and all databases, up. Box plot of SROCC for each quality measures and all databases, down.}
\label{fig:11}       % Give a unique label
%\end{center}
\end{figure*}

The two most widely used criteria to evaluate IQA methods are expressed in terms of Kendall Rank Order Correlation Coefficient (KROCC) and Spearman Rank Order Correlation Coefficient (SROCC). For more details see \cite{98}. Commonly, the better IQA method should present higher SROCC and KROCC. \\
For all databases, we compute the KROCC and SROCC between the MOS and the values computed by the nine tested quality measures to determine if there are statistically significant differences in terms of the performance between the benchmark and the considered metrics.
Table \ref{tab:1} presents the values of KROCC and SROCC for the considered databases and metrics which are visualized as bar chart in Figure \ref{fig:8} and Figure \ref{fig:9} by databases and by quality measures. 
In Figure \ref{fig:10} the bar charts relative to KROCC and SROCC of all quality measures for each dataset are respectively shown. \\
In the following, we perform an evaluation by database and/or by quality measures. To evaluate by database, bar charts for KROCC and SROCC taking as reference the datasets and suitable subparts with respect the quality measures are shown in Figure 8 (left) and Figure 9 (left), respectively. To give evidence to the distribution of the KROCC and SROCC values, the bar charts for KROCC and SROCC of all quality measures for each database are shown respectively in Figure 10 (left) and Figure 10 (right). To evaluate by quality measures, bar charts for KROCC and SROCC taking as reference the quality measures with respect the datasets and suitable subparts are shown in Figure 8 (right) and Figure 9 (right), respectively. The distribution of KROCC and SROCC, for each quality measure on all databases are studied by taking into account the box plot of KROCC and SROCC, respectively shown in Figure \ref{fig:11}.\\
From Table 1, we compute the Quality Measure Ranking (QMR) based on KROCC and SROCC on different set of databases in decreasing order of correlation, shown in Table \ref{tab:2} and Table \ref{tab:3}, where, for short, TIDD*,TIDD*CQD; CQD-Mean;  CQD-Wu, CQD-Octree, CQD-SOM, CQD-kmeans are respectively indicated by: TC*-tot, CQD-M, CQD-W, CQD-O, CQD-S,CQD-k.  Finally, in Table \ref{tab:4} and \ref{tab:5} the statistical performance of the considered quality measures for Main DB is reported.
% For tables use on two column (mettere *)
\begin{table*}[!t]
\begin{center}
\caption{Performance comparison of quality measures on all databases}
\label{tab:1}    % Give a unique label
\begin{tabular}{llllllllllll}
\hline \noalign{\smallskip}
Criteria &Database & PSNR & SSIM&MSSIM& VSNR & VIFP&UQI& NQM& WSNR&SNR \\
\hline\noalign{\smallskip}
   KROCC & TID*& 0,693 & 0,645 & 0,679 & 0,644 &0,632 & 0,469 & 0,632& 0,632& 0,632 \\
%\noalign{\smallskip}
 &TIDD*&0,638& 0,466&0,671&0,665&0,525&0,401&0,662&0,711&0,589\\
%\noalign{\smallskip}
&CQD&0,649&0,673&0,746&0,750&0,776&0,532&0,731&0,738&	0,725\\
 %\noalign{\smallskip}
&TID*CQD&0,495&	0,460&0,467&0,467&0,484&0,357&0,449&0,471&0,466\\
  %\noalign{\smallskip}
& TIDD*CQD&0,638&0,466&0,671&0,665&0,525&0,401&0,662&0,711&0,589\\
 %\noalign{\smallskip}
&CQD -SOM&0,801&0,746&0,771&0,784&0,790&0,542&0,773&0,768&0,738\\
 %\noalign{\smallskip}
&CQD -Median&0,820&0,776&0,765&0,756&0,787&0,508&0,737&0,762&0,751\\
%\noalign{\smallskip}
&CQD -kmeans&0,805&0,727&0,740&0,776&0,779&0,487&0,754&0,767&0,743\\
 %\noalign{\smallskip}
 & CQD -Octree & 0,835 & 0,601 & 0,808 & 0,813 & 0,831 & 0,625 & 0,807 & 0,810 & 0,784\\
  %\noalign{\smallskip}
 & CQD -Wu & 0,806 & 0,764 & 0,784 & 0,804 & 0,808 & 0,535 & 0,786 & 0,782 & 0,741\\
Av. KROCC &     &0,718 & 0,632 & 0,710 & 	0,712	 & 0,694 & 	0,486	 & 0,699 & 	0,722	 & 0,676   \\
 \noalign{\smallskip}
\noalign{\smallskip}
\noalign{\smallskip}
\noalign{\smallskip}
\noalign{\smallskip}
SROCC&TID*&0,880&0,806&0,866&0,836&0,761&0,664&0,826&	0,889&0,836\\
 %\noalign{\smallskip}
 &TIDD* &0,838 &0,651 &	0,869 &0,861 &0,706 &0,588 &0,858 &0,900 &0,797\\
 %\noalign{\smallskip}
&CQD&0,801&0,861&0,918&0,923&0,938&0,724&0,912&0,916&	0,908\\
 %\noalign{\smallskip}
&TIDD*CQD&	0,655&0,622	&0,629&0,632&0,643&0,504&0,616&0,634&0,630\\
  %\noalign{\smallskip}
& TIDD*CQD&0,838&0,651&0,869&0,861&0,706&0,588&0,858&0.900&0,797\\
 %\noalign{\smallskip}
&CQD -SOM&0,950&0,921&0,934&0,944&0,946&0,740&0,938&0,934&0,913\\
   %\noalign{\smallskip}
&CQD -Median&0,961&0,938&0,931&0,926&0,943&0,701&0,916&0,932&0,926\\
   %\noalign{\smallskip}
&CQD -kmeans&0,952&0,909&0,912&0,938&0,942&	0,677&0,928&0,933&0,916\\
   %\noalign{\smallskip}
&CQD -Octree&0,965&0,797	&0,950&	0,954&	0,958&	0,805&	0,948	&0,952&	0,940\\
  %\noalign{\smallskip}
&CQD -Wu&	0,953&	0,929&	0,939&	0,952&	0,955&	0,728&	0,945&	0,941&	0,917\\
  Av. SROCC	&     &	0,879&	0,809&	0,882&	0,883&	0,850&	0,614&	0,875&	0,893&	0,858\\
 \noalign{\smallskip}\hline
\end{tabular}
\end{center}
\end{table*}
% For tables use on two column (mettere *)
\begin{table*} %[ !t]
%\begin{center}
\caption{Quality Measure Ranking (decreasing order) of KROCC, where for short, TIDD*,TIDD*CQD; CQD-Mean;  CQD-Wu, CQD-Octree, CQD-SOM, CQD-kmeans are respectively indicated by: TC*-tot, CQD-M, CQD-W, CQD-O, CQD-S,CQD-k.}
\label{tab:2}    % Give a unique label
\begin{tabular}{lllllllllll}
\hline \noalign{\smallskip}
Database	&PSNR	&SSIM &MSSIM	&VSNR	&VIFP &UQI	&NQM  &WSNR &SNR\\
\hline\noalign{\smallskip}
Main DB	&TID* &CQD	 &CQD	&CQD	&CQD	&CQD	&CQD	&CQD\\
   %\noalign{\smallskip}
&CQD	&TID*	&TID*	 &TC*-tot	&TID* &TID* & TC*-tot & TC*-tot  &TID*\\
 %\noalign{\smallskip}
& TC*-tot	& TC*-tot & TC*-tot	&TID*  & TC*-tot  & TC*-tot  &TID* &TID* & TC*-tot\\
 %\noalign{\smallskip}
&TID*CQD	&TID*CQD	&TID*CQD	&TID*CQD	&TID*CQD	&TID*CQD	&TID*CQD	&TID*CQD	&TID*CQD\\
\hline\noalign{\smallskip}
  %\noalign{\smallskip}
All	&CQD-Octree	&CQD-M	&CQD-O	&CQD-O	&CQD-O	&CQD-O	&CQD-O	&CQD-O	&CQD-O\\
 %\noalign{\smallskip}
&CQD-M	&CQD-W	&CQD-W	&CQD-W	&CQD-W	&CQD-S	&CQD-W	&CQD-W	&CQD-M\\
 %\noalign{\smallskip}
&CQD-W	&CQD-S	&CQD-S	&CQD-S	&CQD-S	&CQD-W	&CQD-S	&CQD-S	&CQD-W\\
%\noalign{\smallskip}
&CQD-k	&CQD-k	&CQD-M	&CQD-k	&CQD-k	&CQD &CQD-k	&CQD-k	&CQD-k\\
 %\noalign{\smallskip}
&CQD-S	&CQD &CQD &CQD-M &CQD-M &CQD-M &CQD-M	&CQD-M &CQD-S\\
  %\noalign{\smallskip}
&TID*  &TID* &CQD-k&CQD &CQD &CQD-k &CQD &CQD &CQD \\
 &CQD &CQD-O&TID* & TC*-tot &TID* &TID* & TC*-tot &TC*-tot &TID*\\
& TC*-tot	 & TC*-tot & TC*-tot &TID* & TC*-tot & TC*-tot &TID* &TID* & TC*-tot\\
&TID*CQD	&TID*CQD	&TID*CQD	&TID*CQD	&TID*CQD	&TID*CQD	&TID*CQD	&TID*CQD	&TID*CQD\\
\hline\noalign{\smallskip}
C\&sub	&CQD-O	&CQD-M	&CQD-O	&CQD-O	&CQD-O	&CQD-O &CQD-O	&CQD-O	&CQD-O\\
	&CQD-M	&CQD-W	&CQD-W	&CQD-W	&CQD-W	&CQD-W	&CQD-W	&CQD-W	&CQD-M\\
	&CQD-W	&CQD-S	&CQD-S	&CQD-S	&CQD-S	&CQD-S	&CQD-S	&CQD-S	&CQD-k\\
	&CQD-k	&CQD-k	&CQD-k	&CQD-k	&CQD-k	&CQD-M	&CQD-k	&CQD-k	&CQD-W\\
	&CQD-S	&CQD	&CQD-M	&CQD-M	&CQD-M	&CQD-k	&CQD-M	&CQD-M	&CQD-S \\
	&CQD &CQD-O	&CQD	&CQD	&CQD	&CQD	&CQD	&CQD	&CQD\\
 \noalign{\smallskip}\hline
\end{tabular}
%\end{center}
\end{table*}
% For tables use on two column (mettere *)
\begin{table*}[!t]
\begin{center}
\caption{Quality Measure Ranking (decreasing order) of SROCC, where for short, TIDD*,TIDD*CQD; CQD-Mean;  CQD-Wu, CQD-Octree, CQD-SOM, CQD-kmeans are respectively indicated by: TC*-tot, CQD-M, CQD-W, CQD-O, CQD-S,CQD-k.}
\label{tab:3}    % Give a unique label
\begin{tabular}{lllllllllll}
\hline \noalign{\smallskip}
Database	&PSNR	&SSIM &MSSIM	&VSNR	&VIFP &UQI	&NQM  &WSNR &SNR\\
\hline\noalign{\smallskip}
 Main DB	&TID*	&CQD	&CQD	&CQD	&CQD	&CQD	&CQD &CQD &CQD\\
& TC*-tot	&TID* &  TC*-tot & TC*-tot &TID* &TID*&TID* & TC*-tot &TID*\\
&CQD & TC*-tot &TID* &TID* & TC*-tot & TC*-tot & TC*-tot &TID* & TC*-tot\\
&TID*CQD	&TID*CQD	&TID*CQD	&TID*CQD	&TID*CQD	&TID*CQD	&TID*CQD	&TID*CQD	&TID*CQD\\
\hline\noalign{\smallskip}
All	&CQD-O	&CQD-M	&CQD-O	&CQD-O	&CQD-O	&CQD-O	&CQD-O	&CQD-O	&CQD-O\\
	&CQD-M	&CQD-W	&CQD-W	&CQD-W	&CQD-W	&CQD-S	&CQD-W	&CQD-W	&CQD-M\\
	&CQD-W	&CQD-S	&CQD-S	&CQD-S &CQD-S  &CQD-W	&CQD-S	&CQD-S	&CQD-W\\
	&CQD-k	&CQD-k	&CQD-M	&CQD-k	&CQD-M	&CQD	&CQD-k	&CQD-M	&CQD-k\\
	&CQD-S	&CQD  &CQD &CQD-M&CQD-k	&CQD-M	&CQD-M	&CQD-k &CQD-S\\
	& TC*-tot	&CQD-O & TC*-tot& TC*-tot	&TID*	&TID* & TC*-tot &TC*-tot &TID*\\
	&CQD	& TC*-tot	&TID*	TID*	& TC*-tot	& TC*-tot	&TID*	&TID*& TC*-tot\\
	&TID*CQD	&TID*CQD	&TID*CQD	&TID*CQD	&TID*CQD	&TID*CQD	&TID*CQD	&TID*CQD	&TID*CQD\\
\hline\noalign{\smallskip}
C\&sub	&CQD-O	&CQD-M	&CQD-O	&CQD-O	&CQD-O	&CQD-O	&CQD-O	&CQD-O	&CQD-O\\
	&CQD-M	&CQD-W	&CQD-W	&CQD-W	&CQD-W	&CQD-W	&CQD-W	&CQD-W	&CQD-M\\
	&CQD-W	&CQD-S &CQD-S	&CQD-S	&CQD-S	&CQD-S	&CQD-S	&CQD-S	&CQD-k\\
	&CQD-k	&CQD-k	&CQD-k&CQD-k	&CQD-k	&CQD-M	&CQD-k	&CQD-k	&CQD-W\\
	&CQD-S	&CQD&CQD-M	&CQD-M	&CQD-M	&CQD-k	&CQD-M	&CQD-M	&CQD-S\\
	&CQD&CQD-O	&CQD&	CQD&	CQD&	CQD&	CQD&	CQD	&CQD\\
 \noalign{\smallskip}\hline
%\caption{Legend: TDD*CQD is indicated in short as  TC*-tot}
\end{tabular}
\end{center}
\end{table*}
% For tables use on two column (mettere *)
\begin{table*}[!t]
\begin{center}
\caption{KROCC statistical performance of the quality measures for Main DB}
\label{tab:4}    % Give a unique label
\begin{tabular}{lllllllllll}
\hline \noalign{\smallskip}
&     &PSNR	&SSIM		&MSSIM	&VSNR	&VIFP	&UQI	&NQM	&WSNR	&SNR\\
\hline\noalign{\smallskip}
  &PSNR	  &-----	  &-1--1	  &--0--	  &--0--	  &-10-1	  &11111	  &--0--	  &-00-0	  &-10-1\\
%\noalign{\smallskip}
  &SSIM	  &-0--0	  &-----	  &-00-0	  &-00-0	  &-00-0	  &1-11-	  &-00--	  &-00-0	-00-0\\
%\noalign{\smallskip}
  &MSSIM	  &--1--	  &-11-1	  &-----	  &-----	  &-1—1	  &11111	  &-----	  &-0--0	  &-1--1\\
 %\noalign{\smallskip}
  &VSNR	  &--1--	  &-11-1	  &-----	  &-----	  &-1--1	  &11111	  &-----	  &-0--0	  &-1--1\\
  %\noalign{\smallskip}
  &VIFP	  &-01-0	  &-11-1	  &-0--0	  &-0--0	  &-----	  &11111	  &-0--0	  &-0--0	  &-----\\
 %\noalign{\smallskip}
  &UQI	  &00000	  &0-00-	  &00000	  &00000	  &00000	  &-----	  &00000	  &00000	  &00000\\
 %\noalign{\smallskip}
  &NQM	  &--1--	  &-11--	  &-----	  &-----	  &-1--1	  &11111	  &-----	  &-0--0	  &-1--1\\
%\noalign{\smallskip}
  &WSNR	  &-11-1	  &-11-1	  &-1--1	  &-1--1	  &-1--1	  &11111	  &-1--1	  &-----	  &-1--1\\
 %\noalign{\smallskip}
  &SNR	  &-01-0	  &-11-1	  &-0--0	  &-0--0	  &-----	  &11111	  &-0--0	  &-0--0	  &-----\\
 \noalign{\smallskip}\hline
\end{tabular}
\end{center}
\end{table*}
% For tables use on two column (mettere *)
\begin{table*}[!t]
\begin{center}
\caption{SROCC statistical performance of the quality measures for Main DB}
\label{tab:5}    % Give a unique label
\begin{tabular}{lllllllllll}
\hline \noalign{\smallskip}
&     &PSNR	&SSIM		&MSSIM	&VSNR	&VIFP	&UQI	&NQM	&WSNR	&SNR\\
\hline\noalign{\smallskip}
&PSNR	&-----	&-1--1	 &--0--	&--0--	&110-1	&11111	&--0--	&-00-0	 &-10-1\\
&SSIM	&-0--0	&-----	&-00-0	&-00-0	&100-0	&11111	&-00-0	&-00-0	&-00-0\\
&MSSIM	&--1--	&-11-1	&-----	&-----	&11--1	&11111	&-----	&-0--0	&-1--1\\
&VSNR	&--1--	&-11-1	&-----	&-----	&11--1	&11111	&-----	&-0--0	&-1--1\\
&VIFP	&001-0	&011-1	&00--0	&00--0	&-----	&11111	&00--0	&00--0	&0----\\
&UQI	&00000	&00000	&00000	&00000	&00000	&00000	&00000	&00000	&00000\\
&NQM	&--1--	&-11-1	&-----	&-----	&11--1	&11111	&-----	&-0--0	&-1--1\\
&WSNR	&-11-1	&-11-1	&-1--1	&-1--1	&11--1	&11111	&-1--1	&-----	&-1--1\\
&SNR	&-01-0	&-11-1	&-0--0	&-0--0	&1----	&11111	&-0--0	&-0--0	&-----\\
 \noalign{\smallskip}\hline
\end{tabular}
\end{center}
\end{table*}
\subsection{Evaluation by databases}
\label{sec4.6}
Table 1, bar charts in Figure 8 (left), 9 (left) ,10 show that KROCC and SROCC distribution for each database follows a similar trend, i.e. for a given quality measure and database if there is a strong correlation in terms of KROCC index then there is a strong correlation in terms of SROCC index. For instance, let consider the distribution of KROCC and SROCC for TID*, shown in the first row of Figure 10. WSNR and PSNR have the maximal value, followed in decreasing order by MSSIM, then followed by SSIM, VSNR, VFP, NQM, SNR, while UQI have the minimum value.\\
Overall the best performing quality measures by databases are WSNR, MSSIM, VSNR since they display a strong correlation in most of all tested databases, the worst performing quality measure is UQI since it has a weak correlation in all databases, while the remaining performing quality measures can be considered as at an intermediate level. It is interesting to note that this trend, although it appears more flattened, is maintained also on the integrated databases. The presence of distribution flattening is not surprising as the databases have been merged a posteriori and therefore have a high degree of variability.\\
On the contrary, the similarity between the evolution of distributions even on merged databases gives greater support to the experimental data obtained on the other databases, thus allowing to say that the best performing quality measures are WSNR, MSSIM, VSNR, while UQI displays a weak correlation. This suggests that there is no advantage of using such measure UQI in the tested data, and favors the nomination of WSNR, MSSIM, VSNR as metrics that are profitable for CQ evaluation.
\subsection{Evaluation by quality measures}
\label{sec4.7}
Table 1, Table 2, and bar charts in Figure 8 (right) and Figure 9 (right)  show that KROCC and SROCC distribution for each quality measure follows quite the same trend, i. e. for a given quality measure if there is a strong/weak correlation in terms of KROCC index then there is a strong/weak correlation in terms of SROCC index for a given database, while there are little differences in the intermediate values. In particular, the distributions are shown in Figure 8 (right) and Figure 9 (right) reveals that the best performing database is mainly CQD while the worst performing database is TID*CQD, with an intermediate value of TIDD*, TIDD*CQD followed by TID* (or TID* followed by TID*,  TIDD*CQD).\\
Overall the best performing database is CQD characterized by a strong correlation for quite all quality measures. The worst performing database is TID*CQD with a weak correlation for all quality measures, while the remaining databases can be considered as at an intermediate level. These are plausible results because we expect a maximum correlation for the CQD database, which is obtained directly, while we expect the minimum correlation value for the database obtained by fusion by TID and CQD and characterized by maximum data variability. However, since WSNR, MSSIM, VSNR perform well on CQD and TID*CQD, this confirms the hypothesis of better correlation of these measures than the others and suggests that there is some advantage of using UQI measure in CQ evaluation.\\
Further confirmation can be obtained by evaluating the data collected in surveys of different populations taking into account the percentile positions (quartiles) relative to a quarter (Q1), two quarters (or half) (Q2) and three quarters (Q3) of the population, that is, 25\%, 50\%, and 75\% respectively. Indeed, through these percentile positions, it is possible to represent the distribution of data in a compact and very meaningful manner using a box plot that allows us to get information about the dispersion and asymmetry of each deployment and compare two or more distributions. 
In Figure 11 box plots obtained for the different data distributions for the KROCC and SROCC indexes are shown. About Figure 11  (left) related to KROCC, each distribution is asymmetrical, as it is characterized by a different distance between the median and the first quartile and the median and the third quartile. Each distribution is not normal as it is characterized by lengths of Inter Quartile Range (IQR) and by whisker non-proportional to the areas underlying the normal curve between the various quartiles. For all distributions, the distance Q3 and the median are greater than the distance Q1 and the median. For SSIM e VIFP there is an appreciable statistical dispersion (or variance) as they have a high IQR and whisker with small size. For PSNR, VSNR, UQI e SNR there is a comparable statistical dispersion (or variance) as they have IQR smaller but bottom-whisker of larger size for PSNR, VSNR and SNR and top-whisker of larger size for UQI, while for MSSIM, NQM, and WSNR, the IQR is smaller with whiskers having small or null dimension but some isolated anomalous values for MSSIM, NQM, and WSNR, not included in the whiskers, exist.\\
Similarly, by considering Figure 11  (right) for SROCC, each distribution is asymmetrical, as it is characterized by a different distance between the median and the first quartile and the median and the third quartile. Each distribution is not normal as it is characterized by lengths of the Inter Quartile Range and whiskers that are not proportional to the areas underlying the normal curve between the quartiles. For all distributions, the distance between Q3 and the median is greater than the distance between Q1 and the median.\\
For SSIM and VIFP metrics, there is an appreciable dispersion (or variance) because they have a high IQR and whisker with reduced size; for PSNR, UQI, and SNR there is a comparable dispersion (or variance) as they have a smaller IQR but bottom-whisker with the larger size for PSNR, SNR and bottom/up whisker with appreciable size for UQI; while for MSSIM, VSNR, NQM, and WSNR, the IQR is smaller having small or null dimension, but for MSSIM, VSNR, NQM, and WSNR there are some isolated anomalous values that are not included in whiskers.
Overall, the trend of box plots is therefore very similar for KROCC and SROCC. This confirms the increased reliability of data for MSSIM, NQM, and WSNR and the strong similarity already observed in the previous phase through the study of bar charts and the increased reliability of these quality measures.
\subsection{Evaluation by databases and quality measures}
\label{sec4.8}
To further confirm these judgments, the statistical performance of the quality measures is analyzed for the individual main databases. Table 4 shows the results for KROCC where each entry is a codeword of five symbols. The position of the symbol in the codeword represents the following databases (from left to right): TID*, TIDD*, CQD, TID*CQD,  TIDD*CQD. Each symbol gives the result of KROCC on the dataset represented by the position of the symbols. “1” means that the image quality metric from the row is statistically better than the image quality metric from the column, “0 means that it is statistically worse and “-” means that it is statistically indistinguishable. Table  5 shows the results for SROCC. Since a larger number of “0” in the columns (rows) indicates the best (worst) performance while a larger number of “1” in the columns (rows) indicates the worst (best) performance, we evaluate the performance in terms of statistical significance from Table 4. Thus, as expected, WSNR is statistically the best performing metric followed by the MSSIM, VSNR, and NQM while SSIM and UQI metrics have the worst performance.
\section{Discussion}
\label{Ds}
In summary, the quantitative analysis carried out in this paper allows us to reach the following final considerations regarding the performance of the quality measures under consideration and database integration.\\
Based on the above evaluation by measures, by database and on the analysis of the bar chart and the box plots, the quality measures WSNR, MSSIM, VSNR are the best candidates to be used in the CQ application since they display a strong correlation with MOS in all databases. PSNR, SSIM, VSNR, VIFP, NQM, and SNR display a moderate correlation with MOS on all databases. The lowest correlation is achieved by the quality measure UQI, displaying a weak correlation with MOS in all databases. Additionally, as expected, we have found that the tested quality measures are characterized by statistically significant differences in terms of correlation and achieve a strong correlation in CQD and a weak correlation in TID*CQD.\\
We point out that the good performance of WSNR, MSSIM, and VSNR is probably because these metrics for definition simulate better the HVS properties, although the color distribution is not explicitly considered, while the fact that PSNR, SSIM, VSNR, VIFP, NQM, and SNR have an intermediate level performance is probably due to the global and physical characterization of these metrics in terms of contrast, luminosity, etc. The worst performance of UQI can be explained by the fact that this metric is able mainly to detect structural distortions, while CQ has as its primary effect the reduction and the alteration of the contrast and the color space and as the secondary effect the distortion to the structural level. \\
Finally, these results are compatible in terms of statistical significance with those obtained in \cite{26} and \cite{27}. At the moment, other comparisons are impossible to do since different quality measures or distortions are taken into account in the other available papers.
The experimental values for different quality measures are essentially compatible and maintain a strong similarity both on individual databases and databases obtained by integration. Therefore, the database integration can be validated and it is permissible to consider this set of data to evaluate quality measures and the quantitative performance evaluation on each database can be considered as an indicator for performance on the other databases.
\section{Conclusions}
\label{Cf}
The purpose of this paper is to compensate for the lack of an adequate IQA for CQ a) by assessing specific most popular full reference quality measures in the presence of CQ distortions on appropriate available databases; b) by proposing a new performance  evaluation process  of the considered quality measures based on recommendations for best practices.\\
For this goal, we review and evaluate nine state-of-the-art quality measures on two publicly available and subjectively rated image quality databases for CQ degradation, by considering suitable their combinations or subparts.\\
The experiments show that the evaluation of the statistical performance of the quality measures for CQ is significantly impacted by the selected image quality database but, due to the detected strong similarity, also that the quantitative performance evaluation on each database is an indicator for the performance on the other databases. Moreover, the quality measures WSNR, MSSIM, VSNR have closer performances in terms of their correlation to the subjective human rating, while UQI does not achieve a strong correlation with subjective scores. The detected strong similarity both on individual databases and on databases obtained by integration provides the ability to validate the integration process and to consider the quantitative performance evaluation on each database as an indicator for performance on the other databases. \\
Therefore, an important contribution of the paper is  to address the choice of suitable quality measures for CQ since the results achieved in this work can be considered useful by readers interested in the design and evaluation of CQ methods. Another non-negligible contribution of the paper is to define a new performance evaluation process which can be extended to other sets of quality measures, i. e. color-based quality measures. Also, in the paper, some implications and indications which can be usefully considered for the development of appropriate new metrics for CQ, are given.
As soon as possible, future evaluation should further extend the databases including a larger number of images with CQ distortion as well as other quality measures directly related to color information and HVS properties to generalize the evaluation process and validating the results obtained in this paper.
\begin{acknowledgements}
This work has been supported by the GNCS (Gruppo Nazionale di Calcolo Scientifico) of the INDAM (Istituto Nazionale di Alta Matematica).
\end{acknowledgements}
 \section*{Conflict of interest}
The authors declare that they have no conflict of interest.
\bibliographystyle{spbasic}      % basic style, author-year citations

\end{document}